\documentclass[12pt]{article}
\usepackage[dvips]{graphicx}
\usepackage{amsbsy}
\usepackage{amssymb}
\usepackage{float}
\usepackage{wrapfig}
\usepackage{color}

\newcommand{\beq}{\begin{equation}}
\newcommand{\eeq}{\end{equation}}
\newcommand{\bdm}{\begin{displaymath}}
\newcommand{\edm}{\end{displaymath}}
\newcommand{\beqr}{\begin{eqnarray}}
\newcommand{\eeqr}{\end{eqnarray}}
\newcommand{\beqrn}{\begin{eqnarray*}}
\newcommand{\eeqrn}{\end{eqnarray*}}

\begin{document}

\title{\bf Two species of semi-local cosmic strings \\ in Abelian gauged $\mathbb{CP}^2$-sigma models\\  coupled to gravity }

\author{A. Alonso-Izquierdo$^\dag$, W. Garc\'{\i}a Fuertes$^*$,\\
J. Mateos Guilarte$^{\dag\dag}$}
\date{}
\maketitle
\begin{center}
$^{\dag}${{\it Departamento de Matem\'atica Aplicada, Facultad de Ciencias Agrarias y Ambientales,  Universidad de Salamanca,
E-37008 Salamanca, Spain.}}
\\ \vspace{0.2cm} $^*${{\it Departamento de F\'{\i}sica, Facultad de Ciencias,  Universidad de Oviedo,\\
E-33007 Oviedo, Spain.}}
\\ \vspace{0.2cm}
$^{\dag\dag}${{\it IUFFyM, %Facultad de Ciencias,%
 Universidad de Salamanca,
E-37008 Salamanca, Spain.}}
\vskip1cm
\end{center}

\date{ }

\maketitle

\begin{abstract}
\noindent
We study a semi-local Abelian Higgs sigma model with $\mathbb{CP}^2$-valued scalar fields coupled to gravity. We establish the conditions needed for self-duality in this system and obtain cosmic strings, both of self-dual character and out of the self-dual point, in the reference chart of  $\mathbb{CP}^2$. In any of the other two charts of a minimum atlas, the transition functions break the global $SU(2)$ symmetry to the Abelian subgroup $e^{i \sigma^3 \alpha}$. In this context, using the transition functions a new self-dual structure can be constructed, which defines a second species of cosmic strings. Because of the presence of a potential energy for the scalar fields, each species takes values in a different bounded region of the target space and is most naturally described using a particular chart of $\mathbb{CP}^2$.  
\end{abstract}
%\bigskip

%{\bf PACS:}  02.20.Qs,   02.30.Ik, 03.65.Fd.

%\medskip

%{\bf Keywords:} Lie algebras, representation theory,  weight multiplicities
\vfill\eject
\section{Introduction}
Topological defects \cite{vishe94} are long lasting solutions of spontaneously broken field theories whose stability is due to the existence of topologically disconnected sectors in the field configuration space, a property inherited from the
non-triviality of certain homotopy group of the vacuum manifold $\cal{V}$, depending on the dimension of the space and the symmetry of the defect.
In many important cases, e.g. GUTs, the vacuum manifold is an homogeneous coset space ${\cal V}=\frac{G}{H}$ where $G$ is the symmetry group of the system and $H$ the subgroup of $G$ surviving symmetry breaking. Prominent among these topological defects are strings or vortices in four-dimensional space-time that grow themselves along a spacelike curve. Thus the sources of string defects are point defects arising when $\Pi_1({\cal V})\neq 0$ and according to the exact homotopy sequence are classified by $\Pi_0(H)$. The core of the string is a filament of false vacuum which carries a mass per unit length $\mu$ of order $v^2$, where $v$ is the scale of symmetry breaking which, for grand unified theories, is of order $10^{16}$ GeV. Thus, cosmic strings formed during the GUT phase transition are highly massive, $\mu\simeq 10^{22}$ g/cm, and they have a major influence on cosmology and galaxy formation \cite{zel80}. Given that the great mass density is accompanied by a similar tension, curved parts of the string rapidly contract and disappear, so that, finally, the strings settle themselves in straight line configurations. In principle, the existence of string solutions is possible in theories with only global symmetries, although in this case the Goldstone mode which comes together with symmetry breaking makes the mass of the string diverge logarithmically with the distance to the center. If the symmetry is local, however, the gauge fields neutralize the Goldstone mode and give rise to strings with finite mass per unit length, and whose core, appart of false vacuum, contains also trapped magnetic flux. Gravitational effects are also very different in the global and local cases. In the former, there is a gravitational force which decreases inversely with the distance from the center, while in the latter the gravitational field far from the string is effectively null. This is due to the exponential decay of the fields with distance, and to the fact that the string enjoys both translational and boost invariance along its direction, implying that the gravitational pulls of mass density and tension exactly cancel. There is only one remnant in the form of a deficit angle, such that the string appears from long distance as a conic singularity in space. The most important physical effects revealing the presence of a cosmic string are thus gravitational lensing, discontinuities in the Doppler effect for particle moving towards the string and the distortions created by the conic geometry in the gravitational or electromagnetic fields of test particles \cite{vishe94}.

The most simple model for understanding the structure of a cosmic string is, appart of the string with Dirac delta profile of \cite{vil81}, the exact solution of Gott \cite{gott85}. In this solution, the plane transverse to the string direction is divided into two regions: the core, with a finite radius and uniform mass density, and the exterior zone, in which the energy-momentum tensor vanishes. Thus, the geometry of this plane changes abruptly at the borderline: the string interior conforms itself as a spherical cap, while the exterior is a flat cone. There is a critical value $\mu_{\rm crit}$ for the mass density such that if $\mu<\mu_{\rm crit}$ the cap is smaller than a hemisphere and the deficit angle of the exterior region is lower than $2\pi$, while for $\mu>\mu_{\rm crit}$ the transition between the spherical and conical geometries occurs above the equator, the deficit angle exceeds $2\pi$ and the exterior cone is inverted like a dunce cap. This latter case corresponds to the so-called supermassive strings and, when it occurs, the exterior solution closes itself at finite distance and the transverse plane never reaches infinity. Instead, the vertex of the dunce cap can be thought of as a new string of zero radius located in front of the original one. Finally, when $\mu=2\mu_{\rm crit}$, both strings merge, the geometry closes into a complete sphere and there is not a string exterior.

The basic field theory which shows how cosmic strings with finite mass per unit length and a regular profile can arise in the context of realistic particle physics is the Abelian Higgs model \cite{niol73}. This theory describes the interaction between a complex scalar field and a $U(1)$ gauge field and displays an interesting phenomenology. When gravity is disregarded, the model constitutes a relativistic Ginzburg-Landau theory of superconductivity, with the dynamics depending on the balance between two non-dimensional parameters, the gauge coupling $e$ and the Higgs self-coupling $\lambda$. Superconductivity is thus of type I if $\beta=\frac{\lambda}{e^2}<1$ and of type II if $\beta>1$, while in the so-called self-dual limit $\beta=1$ special features such as Bogomolny equations and BPS states emerge. $\Pi_1(\cal{ V})=\mathbb{Z}$ in the AHM and, therefore, the theory supports vortices which are classified by an integer $n$. In multivortex configurations, the forces among vortices are attractive if $\beta<1$ and repulsive when $\beta>1$ \cite{jareb79}, whereas for $\beta=1$ the vortices remain in equilibrium irrespectively of the distances among them. These results indicate that radially symmetric vortices with $n>1$ tend to disaggregate when $\beta>1$ but keep bound otherwise. Once gravity is included in the picture, the solutions become cosmic strings. Cosmic strings in the Abelian Higgs model where discovered by Garfinkle \cite{gar85} and their stability was established by Gregory \cite{gre87}. The self-dual case was subsequently investigated by Linet \cite{lin87, lin88} and Comtet and Gibbons \cite{cogi88}. The AHM cosmic strings with non vanishing cosmological constant where introduced in \cite{li86, ti86} and, in particular, the de Sitter case has been studied in \cite{desitter} and the anti-de Sitter one in \cite{antidesitter}. Along with the strings decaying asymptotically to Minkowski space-time plus a deficit angle, the AHM has also room for cosmic strings approaching a Kasner-Melvin geometry in the exterior region, and also in this case there are the normal and supermassive regimes: for low $\mu$ the length of circles around the string decreases with distance, whereas that when $\mu$ is high enough a circle of infinite perimeter is attained for a finite radius \cite{kasner}. Finally, let us mention that, although in this case on a three-dimensional space-time, the AHM can be replaced by other closely related systems like Maxwell-Chern-Simons-Higgs or pure Chern-Simons-Higgs theories which also harbor gravitating vortices, and these solutions have been also investigated \cite{cs}.

Along the years, the original Abelian Higgs model has been generalized in a variety of forms that conserve however some of its most salient properties, namely the existence of type I and II superconductivies and of a self-dual regime. In particular, it has been considered the inclusion of a dielectric function depending on the scalar field \cite{leenam91}, the incorporation of function multiplying the covariant derivatives, which plays the role of a metric in the scalar field target space \cite{lohe81}, or the combined effect of both factors \cite{ba12}. Also, the non-commutative version of the AHM has been studied, both in its original formulation \cite{lomosch01} or with a dielectric function in place \cite{nosotros14}. These modifications made it possible to ponder variants of the AHM in which the Higgs field is valued on a compact manifold like a sphere \cite{nosotros15, nivi12}, and situations of this type have been also studied in the cosmic string context \cite{vemadlar03}. Another direction in which the Abelian Higgs model has been extended is the addition of more scalar fields, giving rise to the so called semi-local models. The insertion of a new complex scalar field into the AHM  enlarges the symmetry group to a tensor product of a global $SU(2)$ times a gauge $U(1)$ groups, and makes the model portray the limit of the bosonic sector of electroweak theory for Weinberg angle $\theta_W=\frac{\pi}{2}$. The vacuum manifold is in this model simply connected, ${\cal V}=\mathbb{S}^3$, but there are nonetheless vortices, which are dynamically stable in the Type I superconductivity domain, but whose magnetic field tends to disperse into broad lumps when superconductivity changes to type II \cite{vaachu91, hind92, giorrusa92}. In the self-dual regime, there is neutral equilibrium among confined vortices and lumps, and the widening of flux is thus a flat direction in moduli space. Models of this type have been recently studied coupled to gravity \cite{haurr10,halosouurr13}. Here, as in the usual AHM, it is interesting to consider generalized semi-local models with dielectric functions or with scalar fields valued on compact manifolds, although in this case to attain self-duality there is the requirement that the metric in field space has to be K\"{a}hler. It should be noted that generalizations of this kind are well motivated physically and deserve a detailed analysis. In fact, as it is well known, self-duality has its roots in supersymmetry, but a likely role of supersymmetric field models in physics is to be low-energy effective theories describing the dynamics of other, more fundamental, degrees of freedom. From this perspective, while the ordinary semilocal model could be interpreted as the bosonic sector of a fundamental supersymmetric theory, if we allow for non-renormalizable interactions and try to build a low-energy effective theory we are forced to introduce a K\"{a}hler potential and a dielectric function \cite{weinberg00}, and are thus led to the type of generalized systems mentioned above.

In particular, with several scalar fields, the simplest case is a semi-local sigma model with target space $\mathbb{CP}^2$. This model was studied in \cite{nosotros15} focusing on the analysis and description of the two species of vortices found in the system. The central theme of the present article is to extend the treatment given in \cite{nosotros15} to the gravitating case, finding thus $\mathbb{CP}^2$-valued semi-local cosmic strings of two species. We organize the paper as follows. Section 2 defines the model to be studied and fixes the field equations and boundary conditions obeyed by the cosmic strings encompassed in it in the reference chart. In Section 3, self-duality of the model is established and  radially symmetric solutions representing self-dual cosmic strings found. Non self-dual cosmic strings are the subject of Section 4. In Section 5 we shift the playground to the second chart of $\mathbb{CP}^2$ and describe the self-dual cosmic strings arising in a new type of semi-local theory where the global symmetry group is also $U(1)$. Some final comments are offered in Section 6.

Throughout this work we follow the seminal papers \cite{gar85}, \cite{lin87} and \cite{lin88}, and the excellent book \cite{vishe94}, adapting the treatment given there for the gravitating Abelian Higgs Model to the case of the sigma $\mathbb{CP}^2$ theory. Appart from the original article \cite{vaachu91} in which they appeared for the first time, another very useful references to deal with semi-local models in Minkowski space-time are \cite{hind92} and \cite{giorrusa92}.

\section{The semi-local nonlinear $\mathbb{CP}^2$-sigma model coupled to gravity}
We shall deal with a version of the Abelian Higgs model in which  the complex scalar fields are maps from Minkowski space-time to the complex projective space $\mathbb{CP}^2$. This complex manifold is the quotient manifold of $\mathbb{C}^3-\{(0,0,0)\}$ modulo proportionality, i.e. modulo the equivalence relation $(z_1,z_2,z_3)\simeq(z^\prime_1,z^\prime_2,z^\prime_3)$ if $z_a^\prime=w z_a, a=1,2,3$ for some complex number $w\neq 0$. This manifold has complex dimension two and a minimum atlas is formed by three charts $V_1, V_2$ and $V_3$, such that $z_a\neq 0$ in $V_a$ and the chart is equipped with a pair of complex inhomogeneous coordinates by dividing the 3-tuples of each equivalence class by their $z_a$ element. $\mathbb{CP}^2$ is endowed with a K\"{a}hler structure in a natural way such that the metric tensor, the Fubini-Study metric is, in the reference chart with inhomogeneous coordinates $(\psi_1,\psi_2)$, of the form
\bdm
ds^2_{\mathbb{CP}^2}=h_{p\bar{q}}(\psi,\psi^*) d\psi_p d\psi_q^*,\hspace{2cm} h_{p\bar{q}}=\frac{\partial^2 K}{\partial\psi_p\partial\psi_q^*}
\edm
where the K\"{a}hler potential is
\bdm
K(\psi,\psi^*)=4 v^2 \log\left(1+\frac{|\psi_1|^2+|\psi_2|^2}{v^2}\right)\equiv 4 v^2 \log {\cal D} \, .
\edm
Explicitly
\beqrn
h_{1\bar{1}}=\frac{4}{{\cal D}}-\frac{4 |\psi_1|^2}{v^2{\cal D}^2}\hspace{3cm}&&h_{1\bar{2}}=-\frac{4\psi_1^*\psi_2}{v^2{\cal D}^2}\\
h_{2\bar{1}}=-\frac{4\psi_2^*\psi_1}{v^2{\cal D}^2}\hspace{3.5cm}&&h_{2\bar{2}}=\frac{4}{{\cal D}}-\frac{4 |\psi_2|^2}{v^2{\cal D}^2}.
\eeqrn
Notice that our conventions are such that if we pass from $\mathbb{CP}^2$  to $\mathbb{CP}^1\simeq \mathbb{S}^2$ by taking one of the two inhomogeneous coordinates, say $\psi_2$, to vanish, the Fubini-Study metric reduces to the standard round metric on $\mathbb{S}^2$, with $v$ the radius of the sphere and $\psi_1=v \frac{y_1+ i y_2}{v-y_3}$, where $(y_1,y_2,y_3)$ are real Euclidean coordinates in target space and $y_1^2+y_2^2+y_3^2=v^2$. Since $h_{1\bar{1}}$ and $h_{2\bar{2}}$ are real and $h_{1\bar{2}}=h_{2\bar{1}}^*$, the metric is Hermitian and $h_{p\bar{q}} M_{p\bar{q}}\in \mathbb{R}$ if $M_{p\bar{q}}$ is a Hermitian matrix. Besides, the form of the K\"{a}hler potential makes it obvious that the Fubini-Study metric is isometric under $SU(2)$ transformations of the form $\chi_p=U_{pl} \psi_l$, $U\in SU(2)$.

We next consider physical models in which the inhomogeneous coordinates are promoted to complex scalar fields $\psi_1(x)$ and $\psi_2(x)$ living on curved four-dimensional space-time and taking values on $\mathbb{CP}^2$. The action is of the form
\beq
S=\int d^4 x\sqrt{-g}\cal{L},\label{accion}
\eeq
where we choose a space-time metric $g_{\mu\nu}$ of mostly plus signature and the Lagrangian density is
\beq
{\cal L}=-\frac{1}{4} {\cal E}(|\psi|) F_{\mu\nu} F^{\mu\nu}-\frac{1}{2} h_{p\bar{q}}(\psi,\psi^*) D_\mu\psi_p D^\mu\psi_q^* -\frac{1}{2} W^2(|\psi|).\label{lagrangiano}
\eeq
We include a dielectric function ${\cal E}$ and, for later convenience, we write the scalar potential as the square of a function $W$. Both ${\cal E}$ and $W$ depend only on $|\psi|^2=|\psi_1|^2+|\psi_2|^2$ in order to respect the $SU(2)$ isometry of the target manifold, and the covariant derivative is defined as $D_\mu\psi_p=\partial_\mu\psi_p-i e A_\mu\psi_p$. The Lagrangian thus exhibits an interplay between global $SU(2)$ and gauge $U(1)$ symmetries, fitting into the paradigm of semi-local theories. In the natural system of units where the Planck constant is $2\pi$ and the speed of light in vacuum is $1$ the physical dimensions of fields and parameters in terms of mass $M$ are: $[A_\mu]=[\psi_p]=M$, $[e]=1$, $[W]=M^2$ and $[{\cal E}]=[h_{p\bar{q}}]=1$. The K\"{a}hler potential  in turn has dimension $[K]=M^2$, since  $[{\cal D}]=1$. A remark on notation: for space-time indices we use the standard Einstein convention with upper and lower indices for, respectively, contravariant and covariant tensorial components, but for internal $\mathbb{CP}^2$ indices we think preferable to keep all indices down in order to reduce cluttering. Repeated internal indices are summed; nevertheless, in some formulas we will write explicitly summation symbols if it is convenient for clarity.

The Euler-Lagrange equations coming from (\ref{accion})-(\ref{lagrangiano}) are
\beqr
\frac{1}{\sqrt{-g}}\partial_\mu\left(\sqrt{-g}{\cal E}(|\psi|) F^{\mu\nu}\right)&=&\frac{i}{2}e\, h_{p\bar{q}}\,\left(D^\nu\psi_p\psi_q^*-\psi_p D^\nu\psi_q^*\right)\label{eq1}\\
\frac{1}{\sqrt{-g}}D_\mu\left(\sqrt{-g} h_{p\bar{q}}\,D^\mu\psi_p\right)&=&\frac{1}{2}\frac{\partial{\cal E}(|\psi|)}{\partial\psi_q^*}F_{\alpha\beta} F^{\alpha\beta}+\frac{\partial h_{p\bar{l}}}{\partial\psi_q^*}D_\lambda\psi_p D^\lambda\psi_l^*+2 W(|\psi|)\frac{\partial W(|\psi|)}{\partial\psi_q^*}\label{eq2}
\eeqr
and must be solved in conjunction with the Einstein equations
\beq
G_{\mu\nu}\equiv R_{\mu\nu}-\frac{1}{2} g_{\mu\nu} R=8\pi G T_{\mu\nu}.\label{einstein}
\eeq
The energy-momentum tensor $T_{\mu\nu}=g_{\mu\nu}{\cal L}-2 \frac{\partial{\cal L}}{\partial g^{\mu\nu}}$ takes the manifestly real and symmetric form
\beqr
T_{\mu\nu}&=&{\cal E}(|\psi|) F_{\mu\lambda} F_\nu^{\;\lambda}-\frac{1}{4}{\cal E}(|\psi|) g_{\mu\nu} F_{\alpha\beta} F^{\alpha\beta}-\frac{1}{2} g_{\mu\nu} W^2(|\psi|)\nonumber\\&+&\frac{1}{2}\left(h_{p\bar{q}}D_\mu\psi_p D_\nu\psi_q^*+h_{q\bar{p}}D_\nu\psi_q D_\mu\psi_p^*-g_{\mu\nu} h_{p\bar{q}} D_\alpha\psi_p D^\alpha\psi_q^*\right)\label{tmunu}.
\eeqr
We are interested in configurations independent of the time $x^0\equiv t$ and of the third spatial coordinate $x^3\equiv z$, and whose energy per unit length along the $z$ axis is finite. We choose thus the Weyl  $A_0=0$ and axial $A_3=0$ gauges, with the consequence that the corresponding covariant derivatives and gauge field strengths identically vanish: $D_t\psi_p=D_z\psi_p=F_{tz}=F_{tk}=F_{zk}=0$, where we denote indices 0 and 3 directly  as $t$ and $z$  and use Latin indices $i,j,k\dots$ for the coordinates $x^1$ and $x^2$ in the plane perpendicular to the $z$ axis.  Accordingly with the situation that we are going to study, we will take for the metric the ansatz, \cite{lin88},\cite{vishe94},
\beq
ds^2=e^{A(x^1, x^2)}(-dt^2+dz^2)+\gamma_{ij}(x^1, x^2) dx^idx^j\label{metric}
\eeq
with all coefficients depending only on the coordinates on the normal plane to $z$. This form of the metric gives a block-diagonal Einstein tensor with $G_{tz}=G_{tk}=G_{zk}=0$ and requires boost invariance in the $t-z$ plane, $G_{tt}=-G_{zz}$. This is consistent with the Einstein equations (\ref{einstein}): using $g_{tk}=g_{zk}=g_{tz}=0$ in (\ref{tmunu}) we see that $T_{tz}=T_{tk}=T_{zk}=0$, whereas the non-vanishing elements are
\beqr
T_{tt}&=&-T_{zz}=e^A\left[ \frac{1}{4} {\cal E}(|\psi|) F_{ij} F^{ij}+\frac{1}{2} h_{p\bar{q}}(\psi,\psi^*) D_k\psi_p D^k\psi_q^* +\frac{1}{2} W^2(|\psi|)\right] \label{Ttt}\\
T_{ij}&=&{\cal E}(|\psi|) F_{ik} F_j^{\;k}-\frac{1}{4}{\cal E}(|\psi|) \gamma_{ij} F_{lm} F^{lm}-\frac{1}{2} \gamma_{ij} W^2(|\psi|)\nonumber\\&+&\frac{1}{2}\left(h_{p\bar{q}}D_i\psi_p D_j\psi_q^*+h_{q\bar{p}}D_j\psi_q D_i\psi_p^*-\gamma_{ij} h_{p\bar{q}} D_k\psi_p D^k\psi_q^*\right).\label{Tij}
\eeqr
Thus, the action per unit length along the $z$-axis for configurations independent of $t$ and $z$ is
\bdm
S=\int dt d^2x e^A\sqrt{\gamma} {\cal L}=-\int dt d^2x \sqrt{\gamma}\, T_{tt}
\edm
and both members of the Euler-Lagrange equations (\ref{eq1}) with $\nu=t$ or $\nu=z$ vanish identically, while in the equation (\ref{eq2}) for the scalar field there are only contributions coming from space-time indices 1 and 2. An important quantity is the energy per unit length along $z$-axis, which is given by
\beq
\mu=-\int d^2x \sqrt{\gamma}\, T^t_{\;t}=\int d^2x\sqrt{\gamma}\left\{ \frac{1}{4} {\cal E}(|\psi|) F_{ij} F^{ij}+\frac{1}{2} h_{p\bar{q}}(\psi,\psi^*) D_k\psi_p D^k\psi_q^* +\frac{1}{2} W^2(|\psi|)\right\}\label{energia}
\eeq
and coincides with minus the action per unit length and time only if $A=0$.
% \bdm
% {\cal L}=-\frac{1}{4} {\cal E}(|\psi|) F_{ij} F^{ij}-\frac{1}{2} h_{p\bar{q}} D_k\psi_p D^k\psi_q^* -\frac{1}{2} W^2(|\psi|)
% \edm
% \beqr
% \frac{e^{-A}}{\sqrt{\gamma}}\partial_k\left(e^A\sqrt{\gamma}{\cal E}(|\psi|) F^{kj}\right)&=&\frac{i}{2}e h_{p\bar{q}}\left(D^j\psi_p\psi_q^*-\psi_p D^j\psi_q^*\right)\label{eqel1}\\
% \frac{e^{-A}}{\sqrt{\gamma}}D_k\left(e^A\sqrt{\gamma} h_{p\bar{q}}D^k\psi_q\right)&=&\frac{1}{2}\frac{\partial{\cal E}(|\psi|)}{\partial\psi_q^*}F_{ij} F^{ij}+\frac{\partial h_{p\bar{l}}}{\partial\psi_q^*}D_j\psi_p D^j\psi_l^*+2 W(|\psi|)\frac{\partial W(|\psi|)}{\partial\psi_q^*}\label{eqel2}
% \eeqr

We are interested in the solutions of the system which are cylindrically symmetric, the seeds indeed of the cosmic strings living in the model. Cylindrical symmetry makes it convenient to use polar coordinates $x^1=\rho=\sqrt{x^2+y^2}$ and $x^2=\phi={\rm arctan}\frac{y}{x}$  in the perpendicular plane to the $z$-axis and to assume for the fields an ansatz of the form
\beq
e A_\rho=0\hspace{2cm}e A_\phi=n-a(\rho),\hspace{2cm} \psi_p=v f_p(\rho) e^{i\varphi_p(\phi)}\ {\rm (no\  sum)},\label{camposradiales1}
\eeq
with $p=1,2$, and
\beq
\varphi_1(\phi)=n_1\phi\hspace{2cm}\varphi_2(\phi)=n_2\phi+\omega,\label{camposradiales2}
\eeq
where $n$, $n_1$ and $n_2$ are integers and $f_1$ and $f_2$ are real functions. To fix the behavior of the fields at the center of the string and at infinity, we shall take for granted a fact to be justified later, namely that the vacuum expectation value of the scalar fields is the same energy scale $v$ which determines the K\"{a}hler structure of $\mathbb{CP}^2$, i.e., we shall assume that $W(v)=0$. Also, we make use of the global $SU(2)$ symmetry to choose $\psi_1$ as the field component taking a non-zero vev at infinity, a choice which implies that the vorticity $n_1$ coincides with $n$. With these stipulations, regularity of the fields at $\rho=0$ and finiteness of the energy per unit length of the string require the boundary conditions
\beqr
f_1(0)=0\hspace{2cm}&&f_2(0)=\mathcal{H}_0\delta_{n_2,0}\hspace{2cm}a(0)=n\label{contorno1}\\
f_1(\infty)=1\hspace{2cm}&&f_2(\infty)=0\hspace{2.5cm}a(\infty)=0,\label{contorno2}
\eeqr
where $\mathcal{H}_0\in\mathbb{R}$. To complete the ansatz, we specify the form of the metric as in \cite{gar85}
\beq
A=A(\rho)\hspace{2cm}\gamma_{ij} dx^i dx^j=d\rho^2+H^2(\rho) d\phi^2\label{metricasimetrica} .
\eeq
Therefore, in order to forbid conic singularities at the origin and to ensure that at large distances the space-time is flat, with the coordinate $t$ measuring the proper time of an asymptotic observer at rest, the boundary conditions
\beqr
A^\prime(0)=0\hspace{2cm}&& A(\infty)=0\label{contorno3}\\
H(0)=0\hspace{2cm}&& H^\prime(0)=1\label{contorno4},
\eeqr
ought to be imposed, the prime denoting derivation with respect to $\rho$. Notice that by fixing the condition $A(\infty)=0$ we discard from the beginning solutions leading to the Kasner-Melvin vacuum far from the origin, and focus instead on a Minkowskian asymptotic space with a deficit angle $\Delta\phi$. This deficit angle is measured by the derivative of $H(\rho)$ at infinity according to the formula
\bdm
\frac{\Delta\phi}{2 \pi}=1-H^\prime(\infty).
\edm
Moreover, cylindrically symmetric solutions will carry a magnetic field and a quantized magnetic flux given by
\beq
 B=* F^{tz}=\frac{1}{H e^A}F_{\rho\phi} \quad , \quad
\Phi_M=\int d^2x H e^A B=\frac{2\pi n}{e},\label{flujo}
\eeq
as a consequence of the finiteness of $\mu$. Notice that the quantization of the magnetic flux comes from the quantized vorticity of the vector field, see (\ref{camposradiales1}).

The metric (\ref{metricasimetrica}) brings along a diagonal Einstein tensor with $G_{\rho\phi}=0$, such that consistence with  the Einstein equations (\ref{einstein}) requires that the energy-momentum tensor is also diagonal. We can check that this is the case by plugging the ansatz  (\ref{camposradiales1}), (\ref{metricasimetrica}) into (\ref{Tij}), verifying in this way that the only non-vanishing components of $T^\mu_{\;\;\nu}$ are
\beqrn
T^t_{\;t}&=&-\frac{1}{2}(\varepsilon_a+\varepsilon_f+\varepsilon_{af}+u)\hspace{1.8cm}T^z_{\;z}=-\frac{1}{2}(\varepsilon_a+\varepsilon_f+\varepsilon_{af}+u)\\
T^\rho_{\;\rho}&=&\frac{1}{2}(\varepsilon_a+\varepsilon_f-\varepsilon_{af}-u)\hspace{2cm}
T^\phi_{\;\phi}=\frac{1}{2}(\varepsilon_a-\varepsilon_f+\varepsilon_{af}-u),
\eeqrn
where we split the elements of the components of the energy-momentum tensor in the four terms used in the second paper of reference \cite{kasner}, which in our case are
\beqrn
\varepsilon_a&=&\frac{{\cal E}(|\psi|)}{e^2 H^2}\left(\frac{da}{d\rho}\right)^2\hspace{4.1cm}\varepsilon_f=v^2 \sum_{p=1}^2 \sum_{q=1}^2 h_{p\bar{q}}\frac{df_p}{d\rho}\frac{df_q}{d\rho}e^{i(\varphi_p-\varphi_q)}\\
\varepsilon_{af}&=&v^2\sum_{p=1}^2 \sum_{q=1}^2\frac{h_{p\bar{q}}}{H^2}{\cal A}_p {\cal A}_q f_p f_q e^{i(\varphi_p-\varphi_q)}\hspace{1.2cm}u=W^2(|\psi|), \quad\quad {\cal A}_p(\rho)=a(\rho)+n_p-n \, .
\eeqrn
We stress that although ${\cal A}_p$ and $\varphi_p$ carry an internal label, they are in fact $\mathbb{CP}^2$ scalars.

Substitution of the cylindrically symmetric ansatz for the fields and the metric in the Euler-Lagrange equations (\ref{eq1}) and (\ref{eq2}) yields the second-order ODE system
\beq
\frac{d}{d\rho}\left(\frac{e^A}{H} {\cal E}(|\psi|)\frac{da}{d\rho}\right)=\frac{1}{2}e^2 v^2 \sum_{p=1}^2 \sum_{q=1}^2 \frac{e^A}{H}({\cal A}_p+{\cal A}_q)h_{p\bar{q}}f_pf_qe^{i(\varphi_p-\varphi_q)}\label{eqelrad1}
\eeq
\beqr
&&\frac{v}{H e^A}\sum_{p=1}^2\frac{d}{d\rho}\left( H e^A h_{p\bar{q}}\frac{df_p}{d\rho}e^{i\varphi_p}\right)-\frac{v}{H^2}\sum_{p=1}^2h_{p\bar{q}} {\cal A}_p^2 f_p e^{i\varphi_p}-\frac{v^2}{H^2}\sum_{p=1}^2\sum_{l=1}^2\frac{\partial h_{p\bar{q}}}{\partial\psi_l}{\cal A}_p{\cal A}_lf_pf_l e^{i(\varphi_l+\varphi_p)}\nonumber\\&&+\frac{v^2}{H^2}\sum_{p=1}^2\sum_{l=1}^2\frac{\partial h_{p\bar{q}}}{\partial\psi_l^*}{\cal A}_p{\cal A}_lf_pf_l e^{i(-\varphi_l+\varphi_p)}=\frac{1}{e^2 H^2} \frac{\partial{\cal E}(|\psi|)}{\partial\psi_q^*}\left(\frac{da}{d\rho}\right)^2+2 W(|\psi|)\frac{\partial W(|\psi|)}{\partial\psi_q^*}\nonumber\\&&+v^2\sum_{p=1}^2\sum_{l=1}^2\frac{\partial h_{p\bar{l}}}{\partial\psi_q^*}\left[\frac{df_p}{d\rho}\frac{df_l}{d\rho}+\frac{{\cal A}_p{\cal A}_l}{H^2} f_p f_l\right]e^{i(\varphi_p-\varphi_l)}\label{eqelrad2},
\eeqr
whereas regarding the Einstein equations, there is an argument given by Garfinkle for the AHM, which depends only on the regularity of the fields at $\rho=0$ and energy-momentum conservation and it is therefore  valid also for our system, which shows that only two of the three possible ones are independent. The independent equations can be expressed in different forms \cite{vishe94}, and we choose here to write them as
\beqr
\frac{1}{H}\frac{d}{d\rho}\Big[\frac{d H}{d\rho}+\frac{1}{2}\Big(H\frac{d A}{d\rho}\Big)\Big]+\frac{1}{4}\Big(\frac{dA}{d\rho}\Big)^2&=&8\pi G T^t_{\;t}\label{einsteint}\\
\frac{d^2A}{d\rho^2}+\frac{3}{4}\left(\frac{d A}{d\rho}\right)^2&=&8\pi G T^\phi_{\;\phi}.\label{einsteinphi}
\eeqr
In particular, integration in $\rho$ of equation (\ref{einsteint}) gives the relation between the deficit angle and the energy per unit length as \cite{gar85}
\beq
\Delta\phi=8\pi G\mu+\frac{\pi}{2}\int_0^\infty d\rho H\left(\frac{dA}{d\rho}\right)^2.\label{garfinkle}
\eeq
Thus,  the problem of finding semi-local $\mathbb{CP}^2$ cosmic strings consists in solving the second-order ODE system (\ref{eqelrad1}), (\ref{eqelrad2}), (\ref{einsteint}) and (\ref{einsteinphi}) with the boundary conditions (\ref{contorno1}), (\ref{contorno2}), (\ref{contorno3}) and (\ref{contorno4}). We will come back to this system later, but first we study in the next section a self-dual regime which makes the problem easier.

\section{Self-duality in the semi-local $\mathbb{CP}^2$-sigma model coupled to gravity}
\subsection{Self-duality and first-order field equations}
The dynamics of the gravitating semi-local $\mathbb{CP}^2$-sigma model is governed by a system of quite complicated second-order differential equations allowing cosmic strings. Let us now search for a self-dual version of the model which simplifies matters by making room to first-order field ODE equations, to be satisfied not only by cylindrically symmetric configurations. In order to do so, we will assume from the start that $A=0$ in the metric, such that the energy and action per unit length along $z$ are proportional. This assumption simplifies also greatly the Einstein tensor, because it implies that the only non-vanishing Christoffel symbols are those with three Latin indices. Thus, the four-dimensional curvature scalar $R$ coincides with the two-dimensional one $R^{(\gamma)}$
coming from $\gamma_{ij}$, whereas this metric, being two-dimensional, gives a trivial Einstein tensor, namely
\bdm
G_{tt}=-G_{zz}=\frac{1}{2} R^{(\gamma)}\hspace{2cm} G_{tk}=G_{zk}=G_{ij}=0\,
\edm
where $R^{(\gamma)}$ is the scalar curvature in the plane perpendicular to the string. This is consistent with $T_{tt}=-T_{zz}$, which, as we have seen, it is true in our model, but requires also $T_{ij}=0$, the old Poincare stability criterion \cite{Byal}, a fact that will be necessary to check after the self-duality equations will be found. An  important ingredient to establish this first-order ODE system is the Levi-Civita tensor of the $\gamma_{ij}$ metric:
\bdm
\varepsilon_{ij}=\sqrt{\gamma}\;\tilde{\varepsilon}_{ij}\hspace{1cm}\varepsilon^{ij}=\frac{1}{\sqrt{\gamma}}\tilde{\varepsilon}^{ij}\hspace{1cm} \tilde{\varepsilon}_{12}=-\tilde{\varepsilon}_{21}=\tilde{\varepsilon}^{12}=-\tilde{\varepsilon}^{21}=1
\edm
which has some useful properties like
\beqrn
\varepsilon_{ij}\varepsilon^{ik}=\delta_j^k&&\hspace{1.5cm}\varepsilon_{ij}\varepsilon^{ij}=2\\
\varepsilon_{ij}\varepsilon_{kl}\gamma^{jl}=\gamma_{ik}&&\hspace{1cm}\varepsilon^{ij}\varepsilon^{kl}\gamma_{jl}=\gamma^{ik}.
\eeqrn
To proceed, we focus our attention on two  quadratic expressions:
\begin{enumerate}
\item The first one, denoted $X_F^2$, is given by
\bdm
X_F^2=\Big(\sqrt{{\cal E}(|\psi|)}F_{ij}\pm\varepsilon_{ij}W(|\psi|)\Big)\Big(\sqrt{{\cal E}(|\psi|)}F^{ij}\pm\varepsilon^{ij}W(|\psi|)\Big)
\edm
and can be directly expanded into an expression containing two terms appearing in the energy density:
\bdm
X_F^2={\cal E}(|\psi|) F_{ij} F^{ij}+2 W^2(|\psi|)\pm\frac{4}{\sqrt{\gamma}}\sqrt{{\cal E}(|\psi|)} W(|\psi|) F_{12}.
\edm
\item The second one, denoted as $|X_D|^2$, is
\bdm
|X_D|^2=h_{p\bar{q}}\Big(D_i\psi_p\pm i\varepsilon_{ik} D^k\psi_p\Big)\Big(D^i\psi_q^*\mp i\varepsilon^{il} D_l\psi_q^*\Big).
\edm
It may be recast as
\bdm
|X_D|^2=2 h_{p\bar{q}} D_k\psi_p D^k\psi_q^*\mp \frac{4}{\sqrt{\gamma}} {\cal R}\hspace{1cm}{\rm with}\hspace{1cm}{\cal R}=\frac{i}{2} \sqrt{\gamma} h_{p\bar{q}} \varepsilon^{ij} D_i\psi_p D_j\psi_q^*.
\edm
In the modulus-argument form of complex functions $\psi_p=|\psi_p| e^{i\chi_p}$ (no sum) the covariant derivatives read $D_i\psi_p=\psi_p(\partial_i \log|\psi_p|-i e V_{ip})$ (no sum)  with $V_{ip}=A_i-\frac{1}{e}\partial_i\chi_p$. Introduce now the real symmetric matrix $M_{pq}=\frac{4}{\cal D}\left(|\psi_p|^2\delta_{pq}-\frac{|\psi_p|^2|\psi_q|^2}{v^2\cal D}\right)$ (no sum) to find
\bdm
{\cal R}=e\,\sqrt{\gamma}\,\varepsilon^{ij}\, V_{ip}\, S_{jp}\hspace{1cm}{\rm where}\hspace{1cm}S_{jp}=M_{pq} \partial_j \log|\psi_q|=2\partial_j\frac{|\psi_p|^2}{\cal D}.
\edm
This implies
\bdm
{\cal R}=2 e\, \partial_j\left(\sqrt{\gamma} \varepsilon^{ij} V_{ip}\frac{|\psi_p|^2}{\cal D}\right)+2 e\, \frac{|\psi_1|^2+|\psi_2|^2}{\cal D} F_{12}
\edm
allowing to write finally  $|X_D|^2$ as
\bdm
|X_D|^2=2\, h_{p\bar{q}} D_k\psi_p D^k \psi_q^*\mp\frac{8e}{\sqrt{\gamma}}\partial_j\left(\sqrt{\gamma} \varepsilon^{ij} V_{ip} \frac{|\psi_p|^2}{\cal D}\right)\mp\frac{4e}{\sqrt{\gamma}} P(|\psi|) F_{12}
\edm
where
\beq
P(|\psi|)=2\frac{|\psi_1|^2+|\psi_2|^2}{\cal D}.
\eeq
\end{enumerate}
The point of introducing $X_F^2$ and $|X_D|^2$ is that we can use them to write the energy per unit length (\ref{energia}) as
\begin{equation}
\mu=\int d^2 x\sqrt{\gamma}\left\{ \frac{1}{4}X_F^2+\frac{1}{4}|X_D|^2\mp\frac{1}{\sqrt{\gamma}}\sqrt{{\cal E}(|\psi|)} W(|\psi|) F_{12}\pm\frac{2e}{\sqrt{\gamma}}\partial_j\left(\sqrt{\gamma} \varepsilon^{ij} V_{ip}\frac{|\psi_p|^2}{\cal D}\right)\pm\frac{e}{\sqrt{\gamma}}P(|\psi|) F_{12} \right\} \, . \label{enden}
\end{equation}
Therefore, the choice of the \lq \lq superpotential" $W(\vert \psi\vert)$ and the dielectric function $\cal{ E}(\vert \psi\vert)$ such that 
\beq
\sqrt{{\cal E}(|\psi|)} W(|\psi|)=e\left( P(|\psi|)-v^2\right)\label{eleccionP}
\eeq
shows $\mu$  as the sum of two squares plus a topological term proportional to the magnetic flux:
\beq
\mu=\int d^2 x\sqrt{\gamma}\left\{ \frac{1}{4}X_F^2+\frac{1}{4}|X_D|^2\right\}\pm e v^2 \Phi_M.\label{bogmu}
\eeq
The choice (\ref{eleccionP}) fixes the generic dependence of $W(\vert\psi\vert)$ on $\cal{E}(\vert\psi\vert)$:
\beq
U(|\psi|)\equiv\frac{1}{2} W^2(|\psi|)=\frac{e^2 v^4}{2 {\cal E}(|\psi|)}\left(\frac{|\psi_1|^2+\vert\psi_2\vert^2-v^2}{|\psi_1^2+\vert \psi_2\vert^2+v^2}\right)^2 \, , \label{potentialsd}
\eeq
which determines the self-dual character of the system. By this statement we mean that the zeroes of $X_F$ and $X_D$, the solutions of the first-order ODE system
\beqr
F_{12}&=&\pm e v^2 \sqrt{\gamma} \frac{v^2-|\psi_1|^2-|\psi_2|^2}{{\cal E}(|\psi|)(v^2+|\psi_1|^2+|\psi_2|^2)}\label{bog1}\\
D_1\psi_p&\pm& i \sqrt{\gamma} D^2 \psi_p= 0\label{bog2}
\eeqr
are absolute minima of $\mu$. The string tension of any solution of the first-order equations with the appropriate boundary condition is thus
\[
\mu=2\pi\vert n\vert v^2
\]
which coincides with the non gravitating $\mathbb{CP}^2$ string tensions unveiled in Reference \cite{nosotros15} if we set $v^2=\frac{a^2}{2}$ as the 
proportionality between the parameters introduced in the present work and \cite{nosotros15}.

Moreover, it is clear from the form of $U(\vert\psi\vert)$ that the set of its zeroes is the set of constant scalar fields such that $\vert\psi_1^{(v)}\vert^2+\vert\psi_2^{(v)}\vert^2=v^2$. Therefore, the vacuum orbit is $\mathbb{S}^3$ if $\cal{E}(\vert\psi\vert)$ has no poles in $\mathbb{CP}^2$ and the vacuum expectation value of the scalar field is any point in $\mathbb{S}^3$, e.g. $\psi_1^{(v)}=v$, $\psi_2^{(v)}=0$, as announced in the previous Section. A subtle point is hidden in the selection of the K$\ddot{\rm a}$hler metric and the vacuum orbit in terms of a unique parameter. On one hand $g^2=\frac{1}{v^2}$ is the (dimensionful)
coupling constant appearing in the non-linear sigma model. On the other hand, $v$ also sets the energy scale of the symmetry breaking. The dual r$\hat{\rm o}$le of $v$ reveals a very economic structure of the space of parameters.

Configurations of finite energy density tend thus to $\mathbb{S}^3$-vacuum orbit $\vert\psi_1^{(v)}\vert^2+\vert\psi_2^{(v)}\vert^2=v^2$ in the circle at spatial infinity in the plane perpendicular to the string:  e.g. $|\psi_1|=v$, $\vert\psi_2\vert=0$ for $|x|\rightarrow \infty$. As recently mentioned, the flux quantization condition (\ref{flujo}) and the structure of (\ref{bogmu}) imply that the bound
\beq
\mu\geq e v^2 |\Phi_M|=2\pi v^2 |n|\label{bound}
\eeq
is saturated if the Bogomolny equations (\ref{bog1})-(\ref{bog2}) are satisfied.
The upper sign gives $F_{12}>0$ ever that $|\psi_1|^2+|\psi_2|^2<v^2$ and corresponds thus to $\Phi_M>0$, i.e., to positive vorticity $n$. Given the proportionality between $\mu$ and the action per unit length along the $z$ axis, and the fact that solutions of the Bogomolny equations give absolute minima of $\mu$, these solutions are also extrema of the actions and thus solutions of the Euler-Lagrange equations.
It remains to check that solutions of (\ref{bog1})-(\ref{bog2}) have vanishing $T_{ij}$. Let us write $T_{ij}=T_{ij}^{(F)}+T_{ij}^{(D)}$ with $T_{ij}^{(F)}$ and $T_{ij}^{(D)}$, respectively, the first and second lines of (\ref{Tij}). Then, using (\ref{bog1}), one finds
\beqrn
T_{ij}^{(F)}&=&\gamma^{kl}{\cal E}(|\psi|) F_{ik}F_{jl}-\frac{1}{4}\gamma_{ij}\Big({\cal E}(|\psi|)F_{mn} F^{mn}+2 W^2(|\psi|)\Big)\\&=&\gamma^{kl} \varepsilon_{ik}\varepsilon_{jl} W^2(|\psi|))-\frac{1}{4}\gamma_{ij}\Big(\varepsilon_{mn}\varepsilon^{mn} W^2(|\psi|))+2 W^2(|\psi|))\Big)=0.
\eeqrn
By means of (\ref{bog2}) in turn, we  unveil the identity
\bdm
D_k \psi_p D^k\psi_q^*=\pm i \varepsilon^{lm}D_l\psi_p D_m\psi_q^*
\edm
from which we derive the formula
\bdm
D_i \psi_p D_j \psi_q^*+D_j \psi_p D_i \psi_q^*=\pm i \gamma_{ij} \varepsilon^{lm}D_l\psi_p D_m\psi_q^*,
\edm
making clear that $T_{ij}^{(D)}=0$. Thus, the Bogomolny equations and the Einstein equations involving latin indices are compatible. Finally, we must consider together with (\ref{bog1})-(\ref{bog2}) the only remaining Einstein equation:
\beq
R^{(\gamma)}= 16 \pi G T_{tt}\label{einsteinautodual}\, .
\eeq
We express the $tt$-component of the energy-momentum tensor as
\bdm
T_{tt}=\frac{1}{2} h_{p\bar{q}} D_i\psi_p D^i\psi_q^*+\frac{e^2 v^4}{{\cal E}(|\psi_1|^2,\vert\psi_2\vert)}\left(\frac{|\psi_1|^2+\vert \psi_2\vert^2-v^2}{|\psi_1|^2+\vert\psi_2\vert^2+v^2}\right)^2,
\edm
using again the Bogomolny equation. We arrive thus to the complete system of equations for obtaining self-dual cosmic strings in the $\mathbb{C}P^2$-sigma model.

In sum self-duality fixes $W(|\psi|)$ in terms of  ${\cal E}(|\psi|)$, but leaves still freedom to choose the dielectric function at will. In what follows, we will focus on the minimal and most natural choice ${\cal E}(|\psi|)=1$. There are, however, other interesting possibilities. We mention two of them. First, by choosing
\beq
{\cal E}(|\psi|)=\frac{2}{\cal D}=\frac{2 v^2}{v^2+|\psi_1|^2+|\psi_2|^2}\label{diesemi}
\eeq
the Bogomolny equations take the form
\beqr
F_{12}&=&\pm \frac{e}{2} \sqrt{\gamma} (v^2-|\psi_1|^2-|\psi_2|^2)\label{bogsemi1}\\
D_1\psi_p&\pm& i \sqrt{\gamma} D^2 \psi_p=0\label{bogsemi2}
\eeqr
which is exactly that of the Bogomolny equations of the standard gravitating semi-local model, even though we are now working with scalar fields taking values on $\mathbb{CP}^2$ rather than in $\mathbb{C}$. The difference is encoded in  the Einstein equation (\ref{einsteinautodual}), where the kinetic term in the $tt$ component of the energy-momentum tensor includes the metric of $\mathbb{CP}^2$, and thus the solutions of (\ref{bogsemi1})-(\ref{bogsemi2}) curve the space-time manifold in a different way than standard semi-local cosmic strings would do.

A second example is
\beq
{\cal E}(|\psi|)=\frac{\mu^2}{4 e^2} \frac{{\cal D}^2}{|\psi_1|^2+|\psi_2|^2}=\frac{\mu^2}{4 e^2 v^4} \frac{(|\psi_1|^2+|\psi_2|^2+v^2)^2}{|\psi_1|^2+|\psi_2|^2}.\label{cs}
\eeq
% which leads to
% \beqr
% F_{12}&=&\pm \frac{e^3}{2\kappa^2} \sqrt{\gamma}(\psi_1|^2+|\psi_2|^2) (v^2-|\psi_1|^2+|\psi_2|^2)\label{bogch1}\\
% D_1\psi_p&\pm& i \sqrt{\gamma} D^2 \psi_p=0\label{bogch2}.
% \eeqr
In the non-gravitational case, and working in 2+1 dimensions, this dielectric function is the appropriate for obtaining self-dual Chern-Simons-Higgs vortices with target $\mathbb{CP}^2$ and Chern-Simons coupling $\mu$. Thus (\ref{cs}) gives self-dual gravitating vortices which converge to Chern-Simons vortices when the Newton constant tends to zero: $G\rightarrow 0$. They are not true Chern-Simons gravitating solitons, however: these carry not only magnetic, but also electric field, and thus angular momentum, which requires a metric with non-diagonal $tk$ components instead of (\ref{metric}). Nevertheless, when $G v^2$ is small,  the r$\hat{\rm o}$le of these non-diagonal components in the solutions of \cite{cs} is subdominant, and the vortices obtained  by means of (\ref{cs}) are good approximations to the solutions of the full Chern-Simons gravitating model. In particular, employing (\ref{cs}) one can extend to the gravitational case some of the phenomenology of Chern-Simons vortices, like the presence of solitons of both topological and non-topological nature whose existence is due to the fact that the vacuum orbit in this case includes also the points $\psi^{(v)}=0$ and $\psi^{(v)}=+\infty$. Models with dielectric functions interpolating between (\ref{diesemi}) and (\ref{cs}) have also been investigated \cite{baz92,nosotros14}.

To end this subsection let us finally compare our system with other very rich non-linear field theory with $\mathbb{CP}^N$ as the target space: the $\mathbb{CP}^N$ model studied, for instance, in the papers \cite{GPe78} or \cite{Wit79} and reviewed in Section 4.5 of \cite{raj82}, where other relevant references can be found. These models are formulated in two-dimensional Euclidean space and include $N+1$ complex scalar fields arranged into a $N+1$ tuple $\vec{n}(x)$ subjected to the constraint $\vec{n}^*(x)\cdot \vec{n}(x)=1$ and the $U(1)$ gauge identification $\vec{n}(x)\equiv \vec{n}\, e^{i \Lambda(x)}$. This system enjoy a very interesting dynamics and display some of the typical features of non-abelian gauge theories, like asymptotic freedom and the presence of instantons. In fact, the model can  be described in terms of an Abelian gauge field minimally coupled to the scalars, and the instantons obey first-order equations, saturate a Bogomolny bound for the action and are classified by a topological number proportional to the magnetic flux of the gauge field. In that sense, there are some remarkable similarities with the theory we are studying, but there are also important differences. In particular, in $\mathbb{CP}^N$ models the gauge field is effective and does not represent any independent degrees of freedom, and thus the Maxwell term is absent\footnote{A Maxwell term and a mass for scalars are generated by radiative corrections in the large $N$ limit. In the effective one-loop theory $U(1)$ is unbroken and there is confinement, but there are not instantons \cite{Wit79}.}. Also, like pure Yang-Mills theory in $(3+1)D$ Minkowski space-time, the $\mathbb{CP}^N$ model is 
scale invariant, so that a potential like the one we have engineered is forbidden. In fact, if we ignore gravity and consider our  $x_1-x_2$ plane as the two-dimensional Euclidean space where a $\mathbb{CP}^2$ model lives, it turns out that the self-duality equations for instantons, when expressed in any
 chart, are simply the Cauchy-Riemann conditions in the complex coordinate $z=x_1+i x_2$ for the fields $\psi_1$ and $\psi_2$. Thus, they coincide with our Bogomolny equations only if the coupling $e$, and thus the potential, vanishes. It seems therefore that our approach and the usual $\mathbb{CP}^N$ models are in some sense complementary: because the scale of instantons ranges from zero to infinity, they disappear from the spectrum in our system.The two ingredients, gauge coupling and potential energy density, required to accomodate cosmic strings or vortices break scale invariance. This complementarity manifests itself also in that regular $\mathbb{CP}^N$-instantons embedded in our model have an infinite energy cost: for instance, the instanton with topological number one is in our coordinates
\beq
\psi_1(z)=\alpha \frac{z-z_0}{\omega},\hspace{2cm}\psi_2(z)=\beta \frac{z-z_0}{\omega},\label{instanton}
\eeq
where $\alpha$ and $\beta$ are complex constants such that $|\alpha|^2+|\beta|^2=1$ and $z_0$ and $\omega$ are, respectively, complex and real arbitrary constants reflecting the translational and scale invariance of the pure $\mathbb{CP}^2$ model. Thus, for $|z|\rightarrow \infty$ the fields diverge in our non scale invariant system: they do not approach the vacuum orbit, the zeroes of the potential, at spatial infinity.
\subsection{Self-dual semi-local $\mathbb{CP}^2$ cosmic strings}
Let us now specialize to cylindrically symmetric solutions. Thus, we take $x^1=\rho$ and $x^2=\phi$ and use the ansatz  (\ref{camposradiales1}), (\ref{camposradiales2}) for the fields and the metric (\ref{metricasimetrica}), working with ${\cal E}(|\psi|)=1$. Plugging these expressions into the Bogomolny equations and the energy-momentum tensor, using that the curvature scalar of (\ref{metricasimetrica}) is $R^{(\gamma)}=-\frac{2}{H}\frac{d^2H}{d\rho^2}$, and turning to non-dimensional quantities by means of the redefinitions
\bdm
x=e v\rho\hspace{1.5cm} h(x)=e v H(\rho)\hspace{1.5cm} 8\pi G v^2=\kappa^2
\edm
we arrive to the ODE system which are the self-duality equations for cylindrically symmetric configuations:
\beqr
\frac{da}{d x}&=&h\frac{f_1^2+f_2^2-1}{f_1^2+f_2^2+1}\label{bograd1}\\
\frac{df_1}{dx}&=&\frac{a}{h} f_1\label{bograd2}\\
\frac{df_2}{dx}&=&\frac{a+l-n}{h} f_2\label{bograd3}\\
\frac{1}{h}\frac{d^2 h}{d x^2}&=&-\kappa^2\left[m_{pq}\frac{df_p}{dx}\frac{df_q}{dx}+\left(\frac{f_1^2+f_2^2-1}{f_1^2+f_2^2+1}\right)^2\right].\label{bograd4}
\eeqr
For definiteness we work with the upper sign of (\ref{bog1})-(\ref{bog2}), we put $n_2=l$, and define $m_{pq}$ as the real symmetric matrix of elements
\bdm
m_{pq}=\frac{4}{1+f_1^2+f_2^2}\left(\delta_{pq}-\frac{f_p f_q}{1+f_1^2+f_2^2}\right).
\edm
\begin{figure}[t]
\centering
\includegraphics[width=5.5cm]{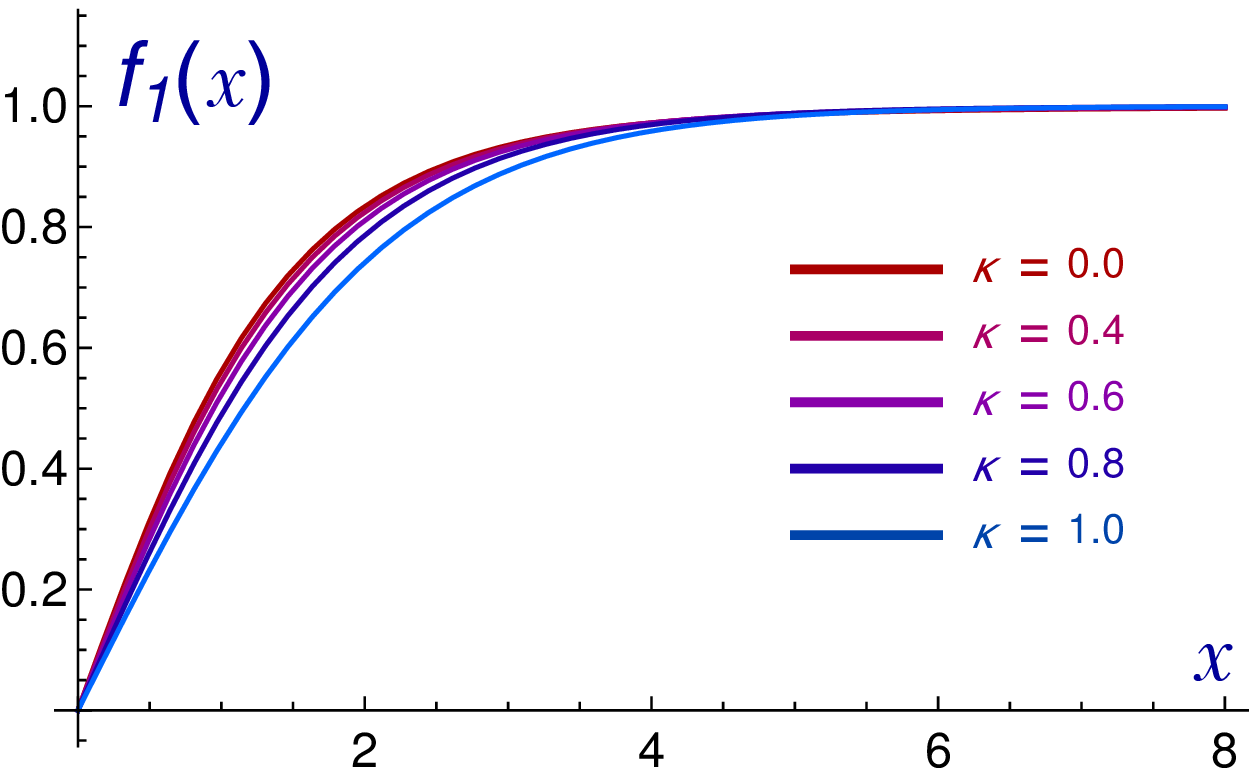}
\hspace{0.2cm}
\includegraphics[width=5.5cm]{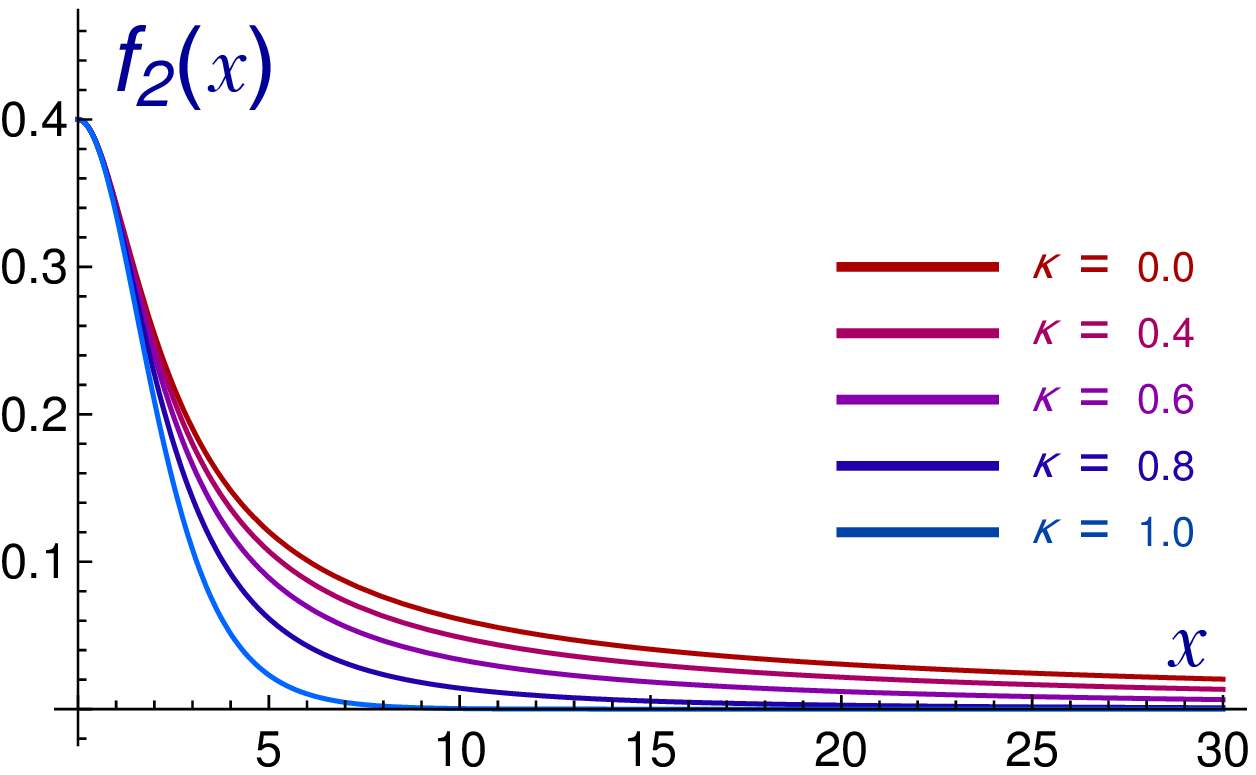}
\hspace{0.2cm}
\includegraphics[width=5.5cm]{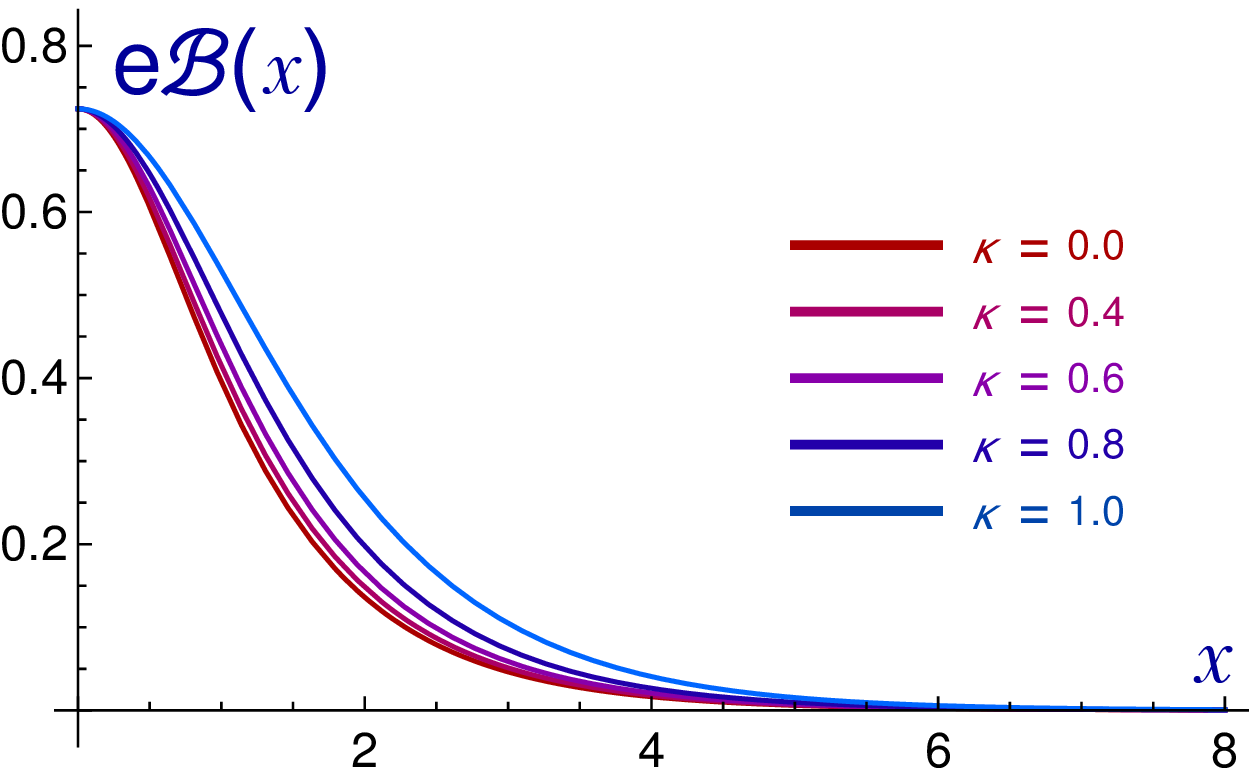}
\caption{Profiles of $f_1$, $f_2$ and the magnetic field for $n=1$ and $\mathcal{H}_0=0.4$ and several strengths of the gravitational interaction.}
\label{fig1}
\end{figure}
\begin{figure}[b]
\centering
\includegraphics[width=5.5cm]{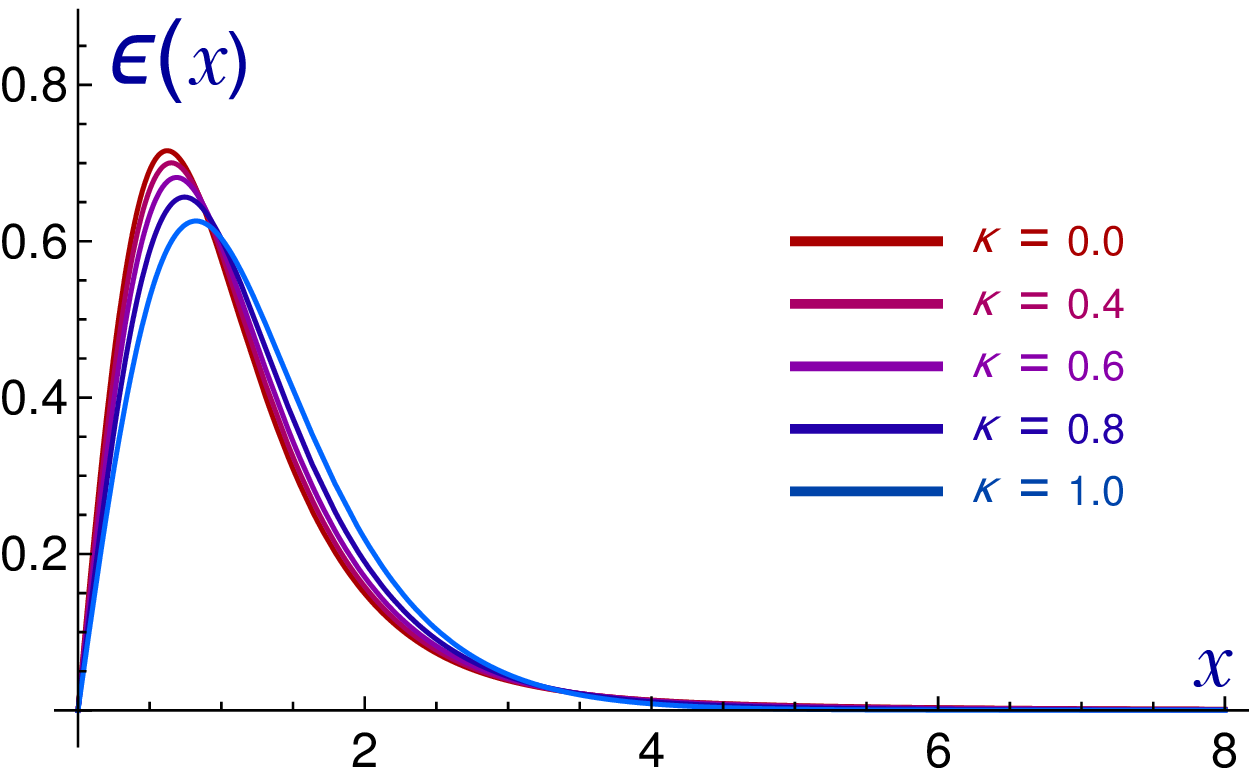}
\hspace{1.5cm}
\includegraphics[width=5.5cm]{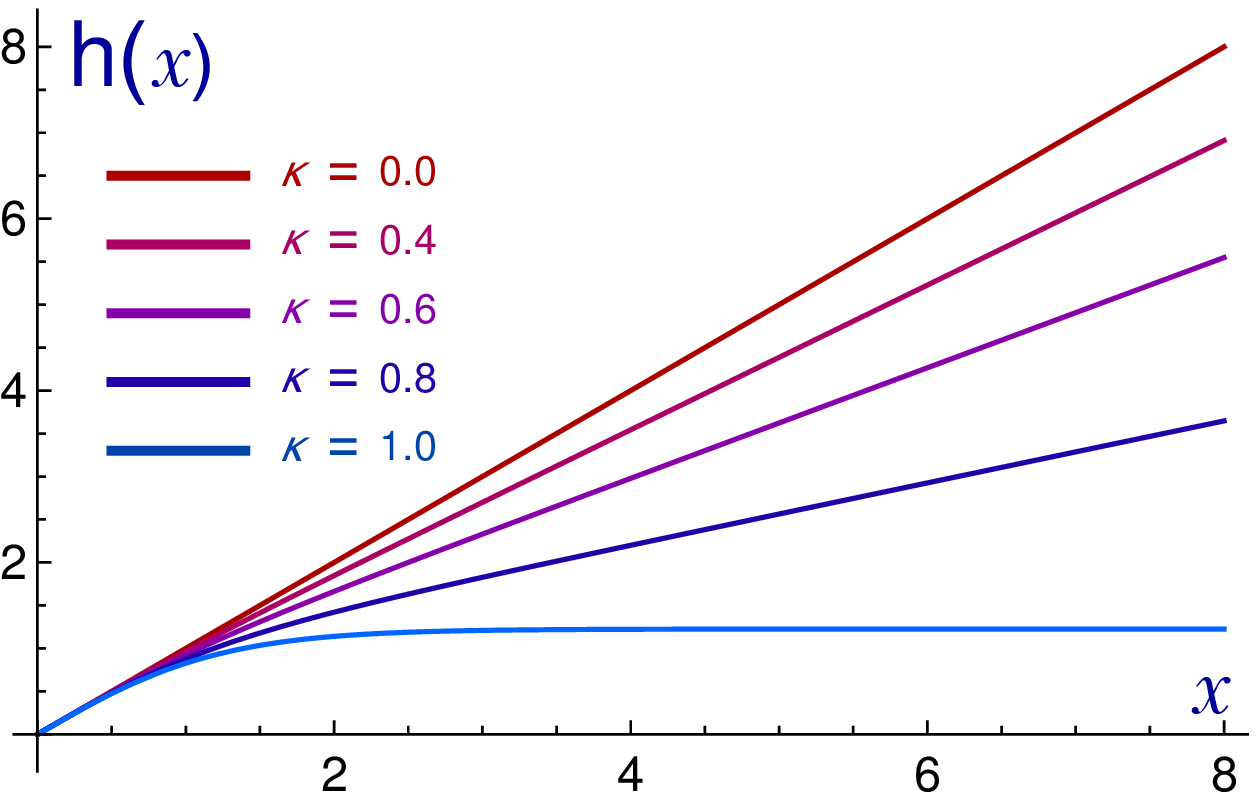}
\caption{Profiles of the energy density and the metric coefficient $h$ for $n=1$ and $\mathcal{H}_0=0.4$ and several strengths of the gravitational interaction.}
\label{fig2}
\end{figure}
The boundary conditions are (\ref{contorno1}), (\ref{contorno2}), (\ref{contorno3}) and (\ref{contorno4}), with the understanding that the prime means now a derivative with respect to $x$. Besides, the non-dimensional magnetic field ${\cal B}=\frac{B}{e^2 v^2}$ and  energy density $\epsilon(x)$, defined by $\mu=2\pi v^2 \int_0^\infty dx\, \epsilon(x)$, are as follows:
\beq
e{\cal B}=-\frac{1}{h}\frac{da}{dx},\hspace{2cm}\epsilon(x)=h(x)\left[m_{pq}\frac{df_p}{dx}\frac{df_q}{dx}+\left(\frac{f_1^2+f_2^2-1}{f_1^2+f_2^2+1}\right)^2\right].
\eeq
\begin{figure}[t]
\centering
\includegraphics[width=5.5cm]{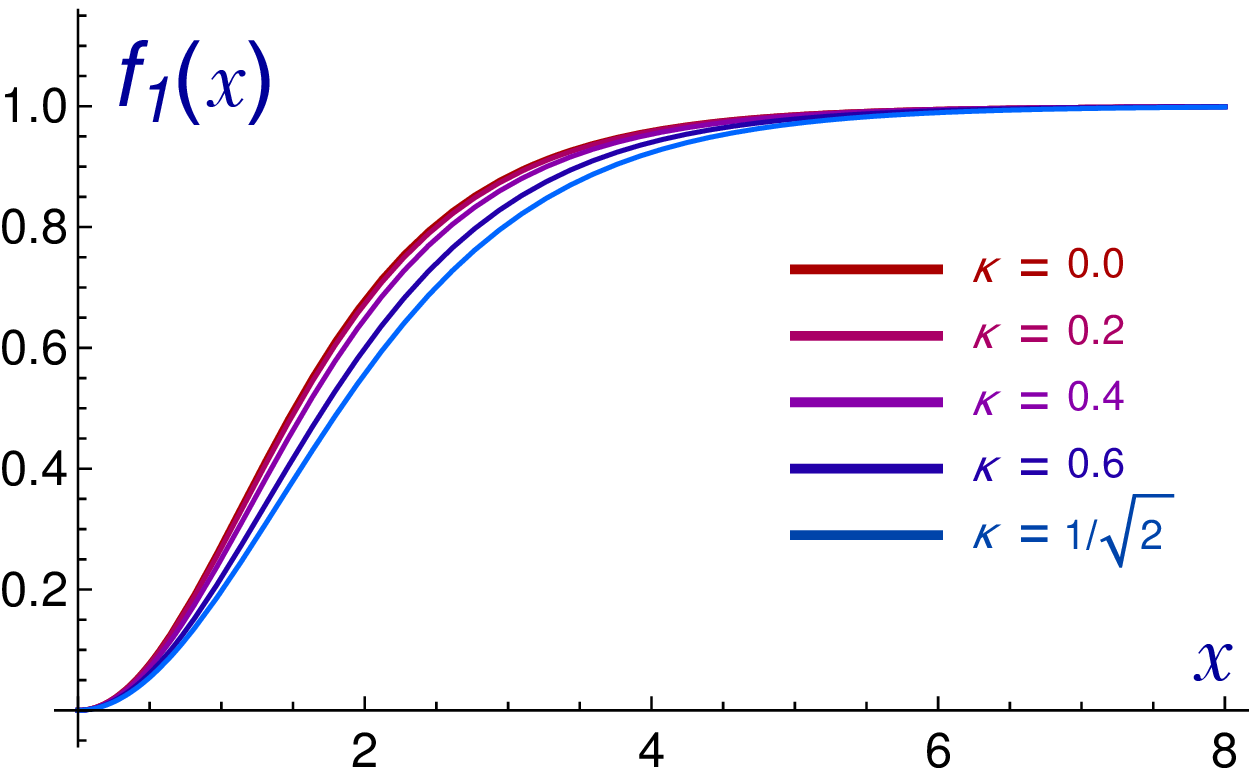}
\hspace{0.2cm}
\includegraphics[width=5.5cm]{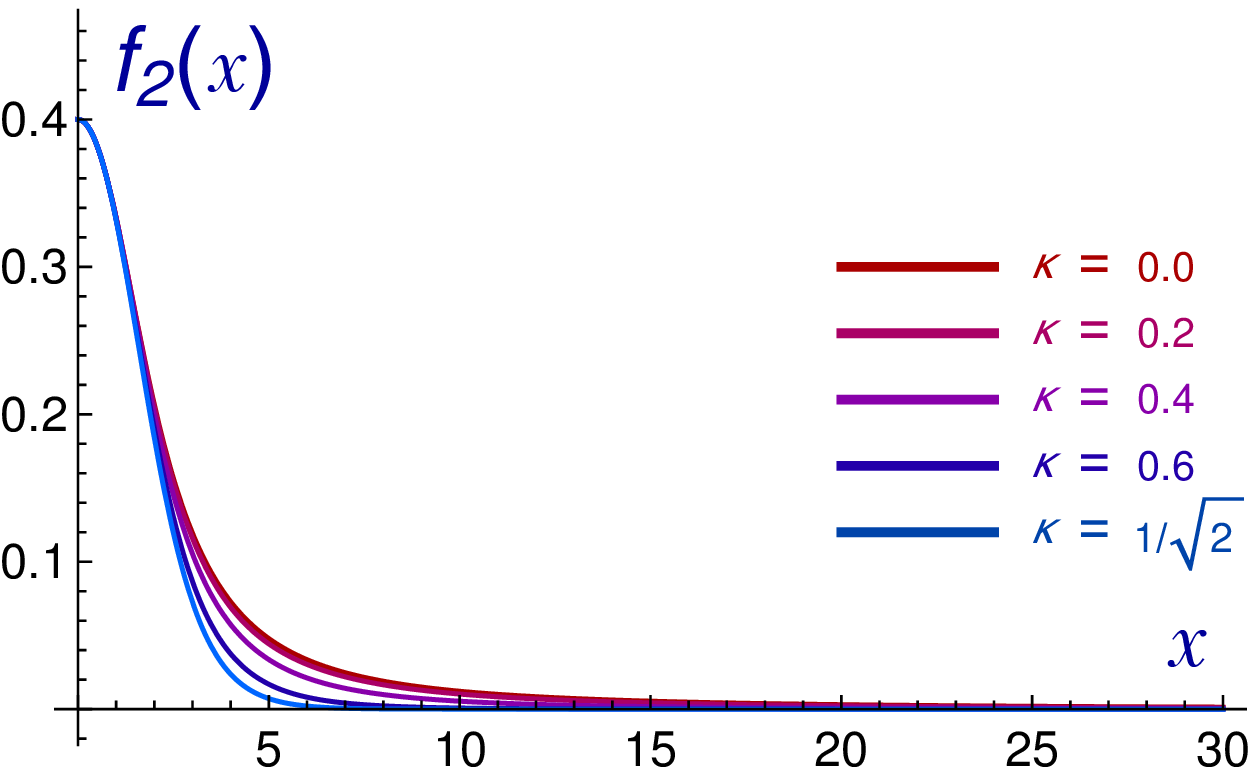}
\hspace{0.2cm}
\includegraphics[width=5.5cm]{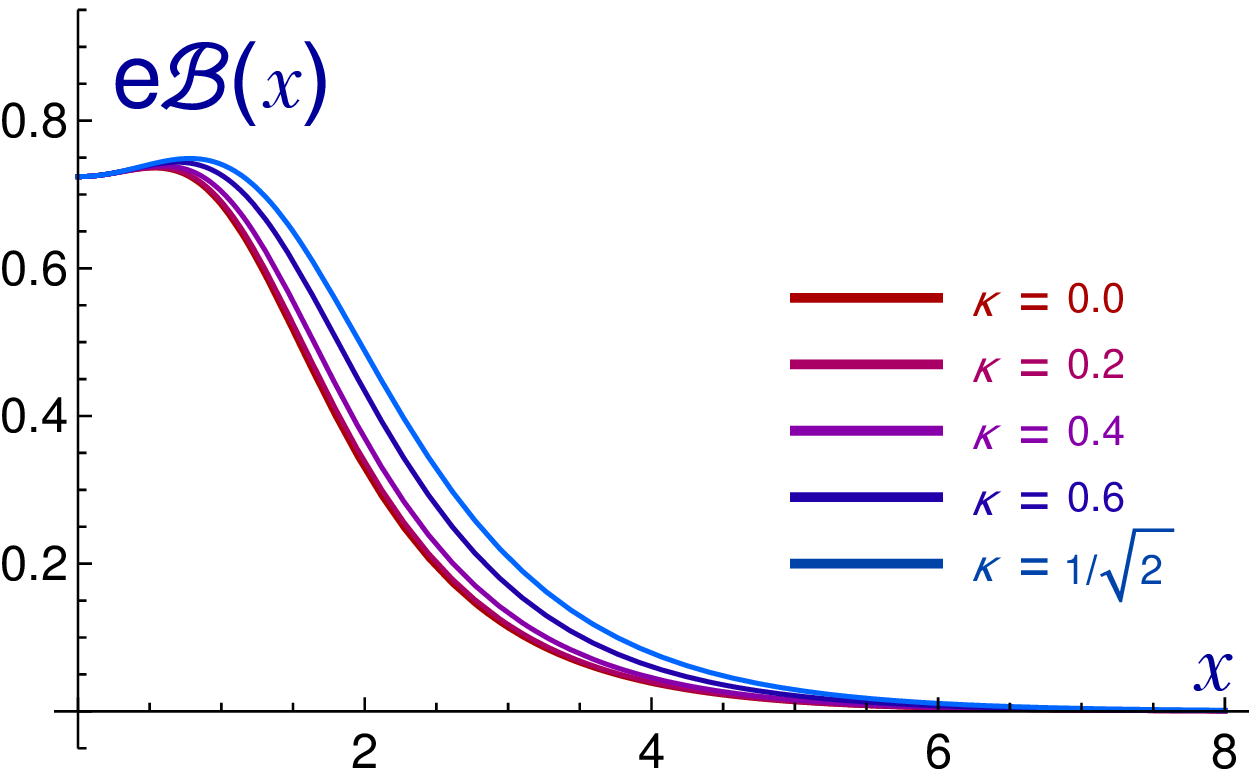}
\caption{Profiles of $f_1$, $f_2$ and the magnetic field for $n=2$, $l=0$ and $\mathcal{H}_0=0.4$ and several strengths of the gravitational interaction.}
\label{fig3}
\end{figure}
The standard procedure in the search of self-dual cosmic string solutions  is a three step shooting approach
\begin{enumerate}
\item First, we solve analytically the system of ODE's (\ref{bograd4}) near the origin. If $x\simeq 0$ we assume the following form of the solutions
\beq
f_1(x)\simeq\alpha_1 x^{r_1},\hspace{1.5cm}f_2(x)\simeq\delta_{l0}\mathcal{H}_0+\alpha_2 x^{r_2},\hspace{1.5cm}a(x)\simeq n-\delta x^s\label{cero}
\eeq
at leading order. Here $\alpha_1$, $\alpha_2$, $r_1$, $r_2$, $\mathcal{H}_0$, $\delta$, and $s$ are constants to be determined by solving the linearized system. For instance, in the case $l=0$ and $\mathcal{H}_0\neq 0$, equation (\ref{bograd2}) implies $r_1=n$, equation (\ref{bograd1}) fixes $s=2$ and $\delta=\frac{1}{2}\frac{1-\mathcal{H}_0^2}{1+\mathcal{H}_0^2}$, whereas (\ref{bograd3}) gives $r_2=2$ and $\alpha_2=-\frac{1}{2}\delta \mathcal{H}_0$. If $l\neq 0$, we find again $r_1=n$  and $s=2$, but now  (\ref{bograd1}) gives $\delta=\frac{1}{2}$ and  (\ref{bograd3}) implies $r_2=l$. In both cases, for $x\simeq 0$, (\ref{bograd4}) implies $h(x)\simeq x$ at leading order. To sum up, for  $x$ near zero, the leading order approximations for the fields and the metric are
\beqr
f_1(x)&\simeq& \alpha_1 x^n\hspace{2.9cm}f_2(x)\simeq \delta_{l0} \mathcal{H}_0\left[1-\frac{1}{4}\frac{1-\mathcal{H}_0^2}{1+\mathcal{H}_0^2} x^2\right]+(1-\delta_{l0}) \alpha_2 x^l\nonumber\\
a(x)&\simeq&n-\frac{1}{2}\frac{1-\delta_{l0}\mathcal{H}_0^2}{1+\delta_{l0}\mathcal{H}_0^2} x^2\hspace{1cm}h(x)\simeq x.\label{x0}
\eeqr
\begin{figure}[t]
\centering
\includegraphics[width=5.5cm]{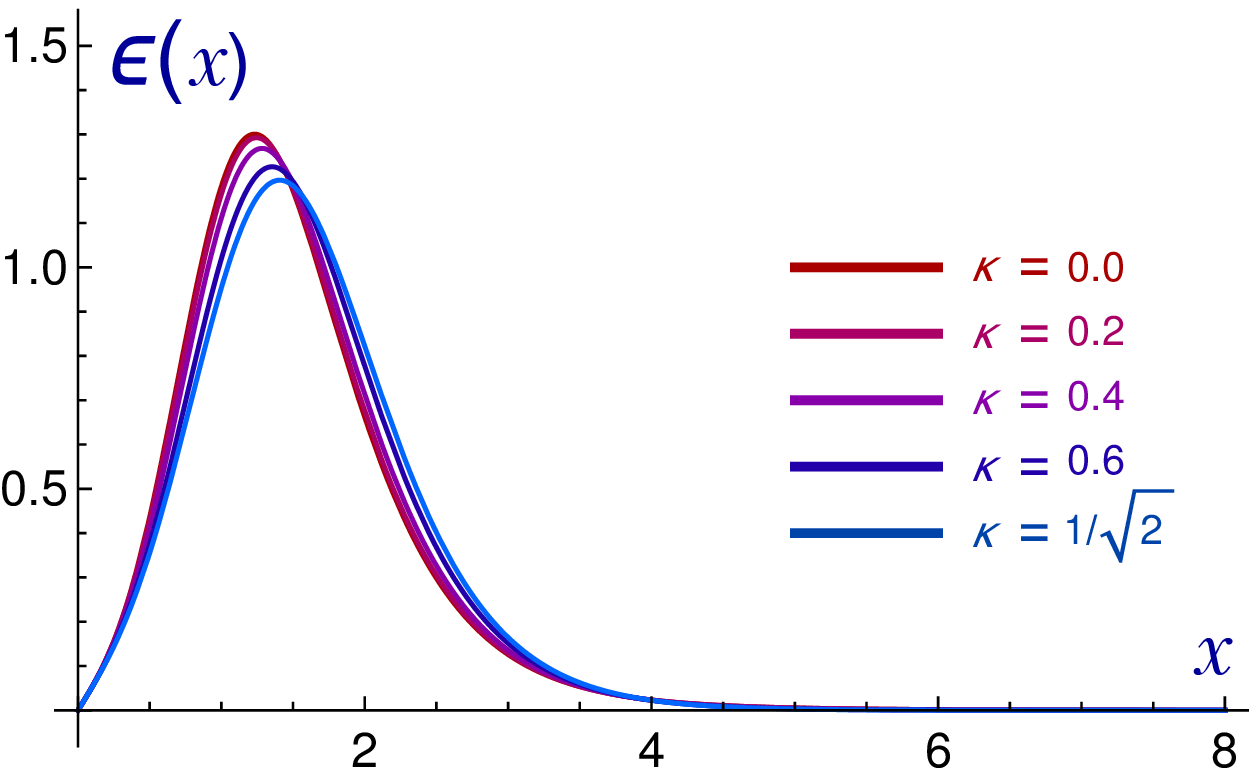}
\hspace{1.5cm}
\includegraphics[width=5.5cm]{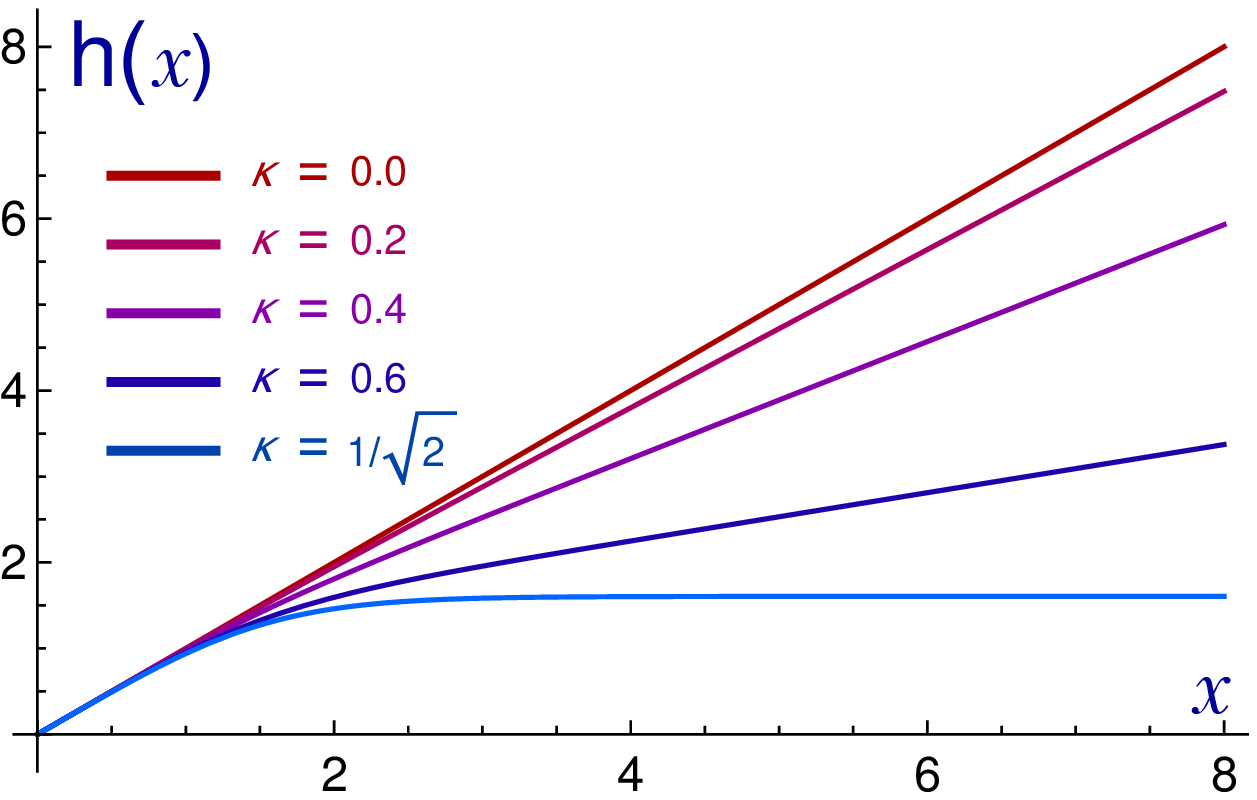}
\caption{Profiles of the energy density and the metric coefficient $h$ for $n=2$, $l=0$ and $\mathcal{H}_0=0.4$ and several strengths of the gravitational interaction.}
\label{fig4}
\end{figure}
\item Second, we solve the system (\ref{bograd4}) for $x\rightarrow\infty$. A first consequence of the boundary conditions at infinity and (\ref{bograd3}) is that $l$ is limited to the values $l=0,1,2,\ldots,n-1$. Given that the energy-momentum vanishes in this region, (\ref{bograd4}) implies a linear profile for $h(x)$ and we approximate the metric very far from the origin in the form $h(x)\simeq c+rx$ where $r$ and $c$ are constants. Then, combining (\ref{bograd2}) and (\ref{bograd3}), it follows that $f_2(x)\simeq C x^\frac{l-n}{r} f_1(x)$ for large $x$ at leading order and some constant $C$. If we put $f_1(x)=1-g(x)$ with $g(x)$ small, from (\ref{bograd2}) $a=-r x \frac{dg}{dx}$ and then from (\ref{bograd1}) we obtain the equation
\bdm
x^2 \frac{d^2 g}{dx^2}+x \frac{d g}{dx}-x^2 g=-\frac{C^2}{2} x^{2\frac{r+l-n}{r}}
\edm
with solution $g(x)=\frac{C^2}{2} x^{2\frac{l-n}{r}}$ for $x\rightarrow \infty$. Thus, very far from the origin, the dominant behavior is
\beq
f_1(x)\simeq 1-\frac{C^2}{2 x^{2\frac{n-l}{r}}}\hspace{1.3cm}f_2(x)\simeq\frac{C}{x^\frac{n-l}{r}}\hspace{1.3cm}
a(x)\simeq (n-l)\frac{C^2}{x^{2\frac{n-l}{r}}}\hspace{1.3cm}h(x)\simeq rx.\label{infinito}
\eeq
% \beqrn
% f_1(x)&=& 1-\frac{C^2}{2 x^{2\frac{n-l}{r}}}\hspace{2.3cm}f_2(x)=\frac{C}{x^\frac{n-l}{r}}\\
% a(x)&=&(n-l)\frac{C^2}{x^{2\frac{n-l}{r}}}\hspace{2.2cm}h(x)= rx.
% \eeqrn
Knowledge of the solution of the cosmic string gravitational field at infinity suffices to determine the deficit angle. Since $A=0$ in the self-dual case, (\ref{garfinkle}) and the saturation of the bound (\ref{bound}) imply $\Delta\phi~=~2\pi\kappa^2 n$, both here and in the standard AHM. Thus, because $r=1-\frac{\Delta\phi}{2\pi}$, this parameter measures the relation  $r=1-\frac{\kappa^2}{\kappa^2_{\rm crit}}$ between the gravitational coupling and the critical coupling $\kappa_{\rm crit}=\frac{1}{\sqrt{n}}$ which is the threshold for the formation of supermassive cosmic strings.
\begin{figure}[t]
\centering
\includegraphics[width=5.5cm]{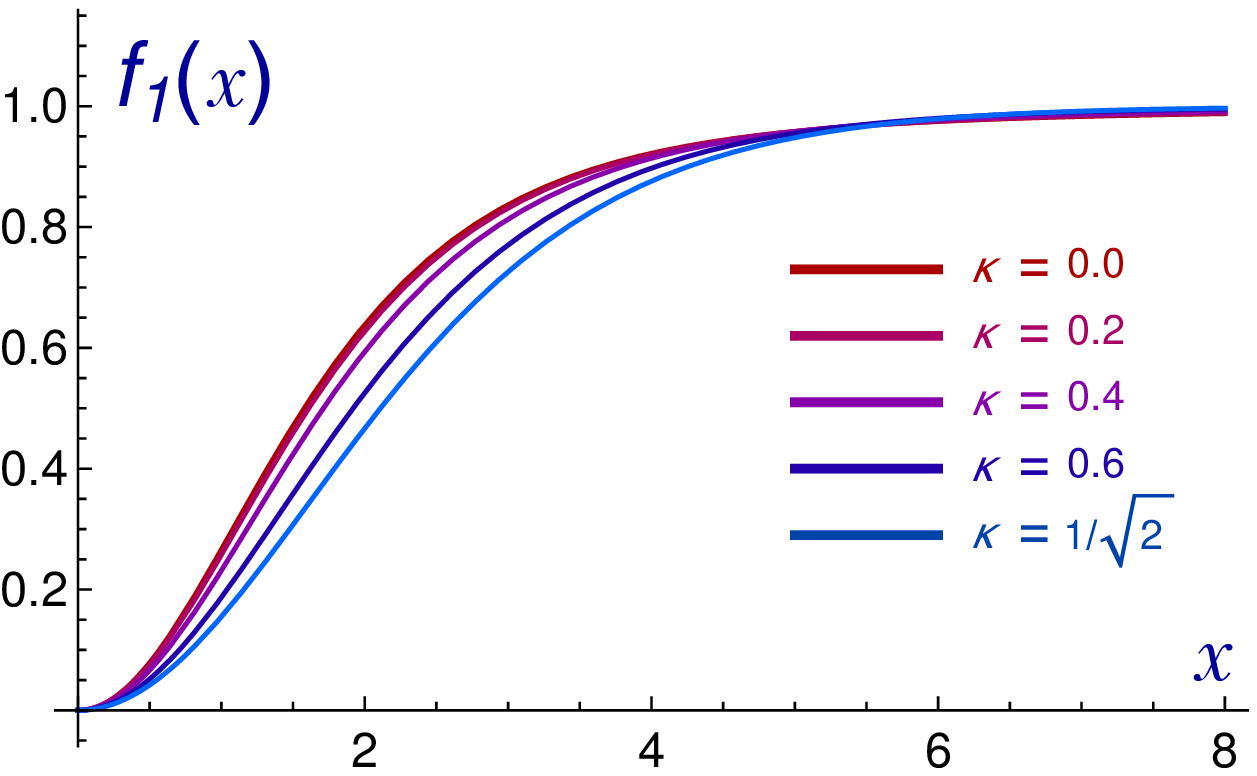}
\hspace{0.2cm}
\includegraphics[width=5.5cm]{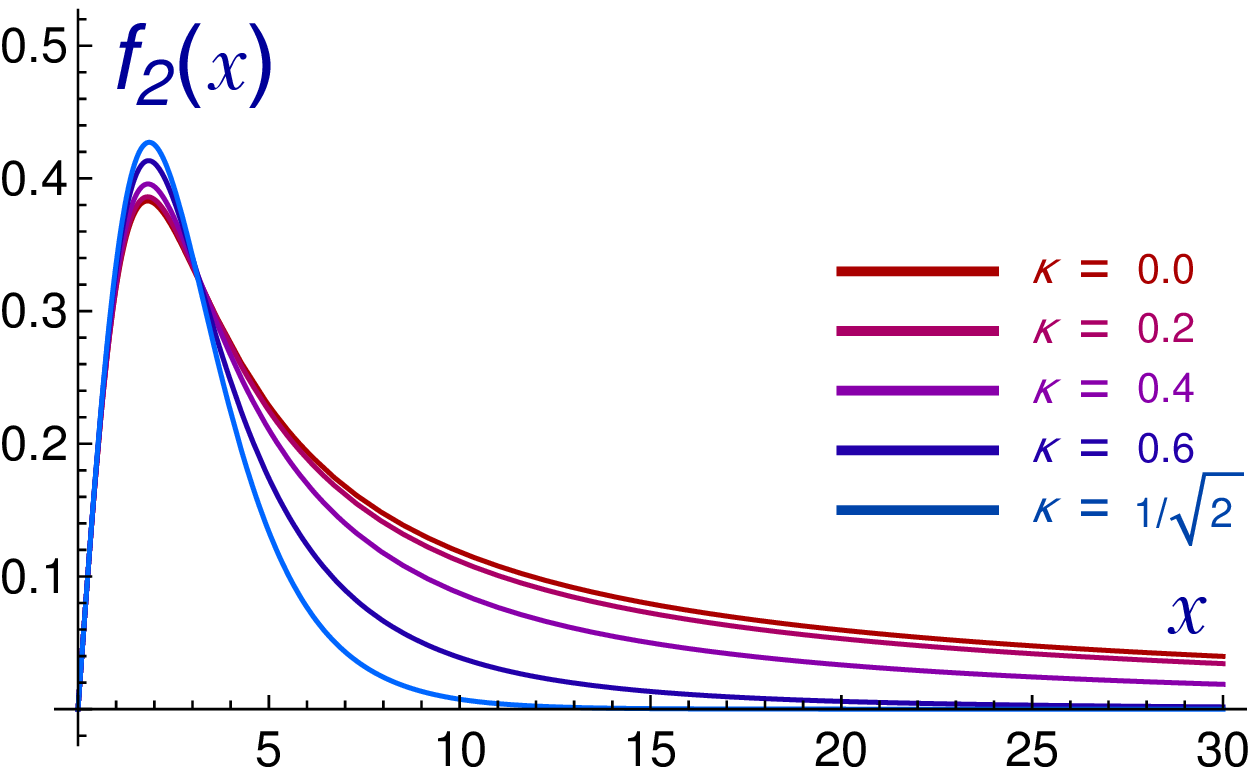}
\hspace{0.2cm}
\includegraphics[width=5.5cm]{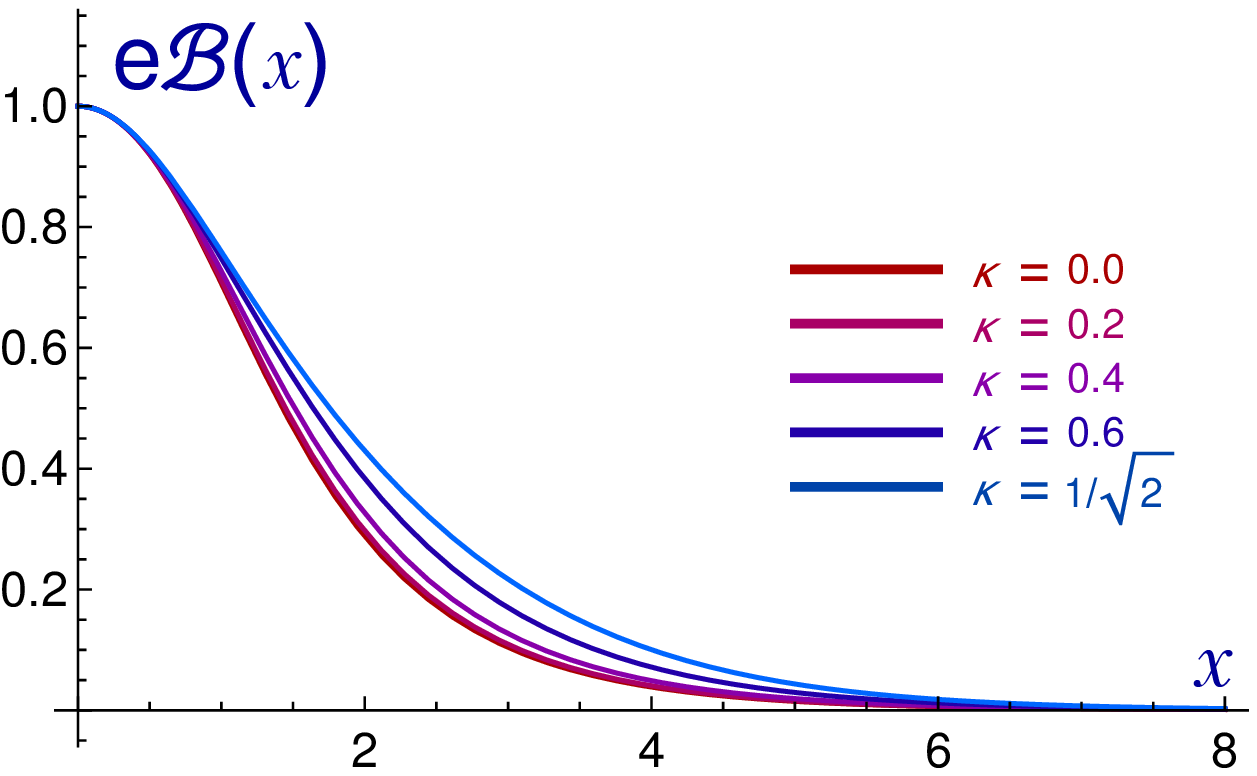}
\caption{Profiles of $f_1$, $f_2$ and the magnetic field for $n=2$, $l=1$ and $\alpha_2=0.4$ and several strengths of the gravitational interaction.}
\label{fig5}
\end{figure}
\begin{figure}[t]
\centering
\includegraphics[width=5.5cm]{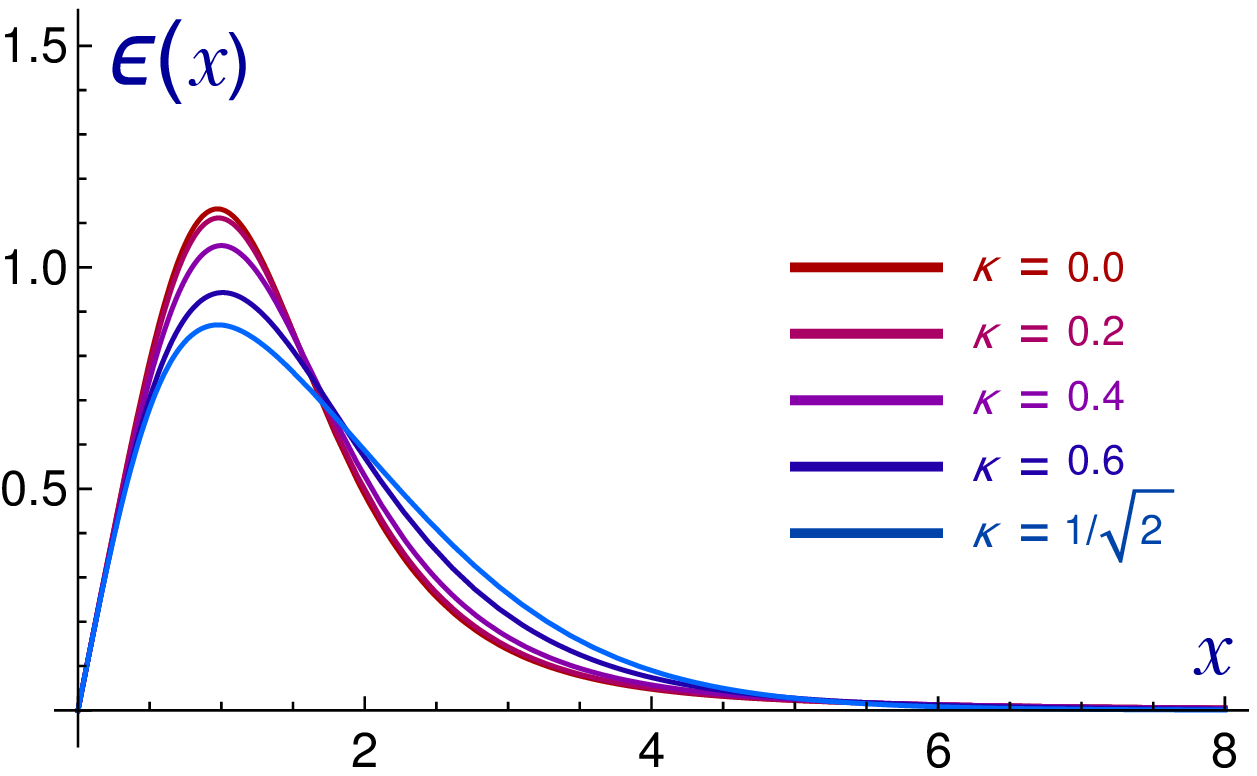}
\hspace{1.5cm}
\includegraphics[width=5.5cm]{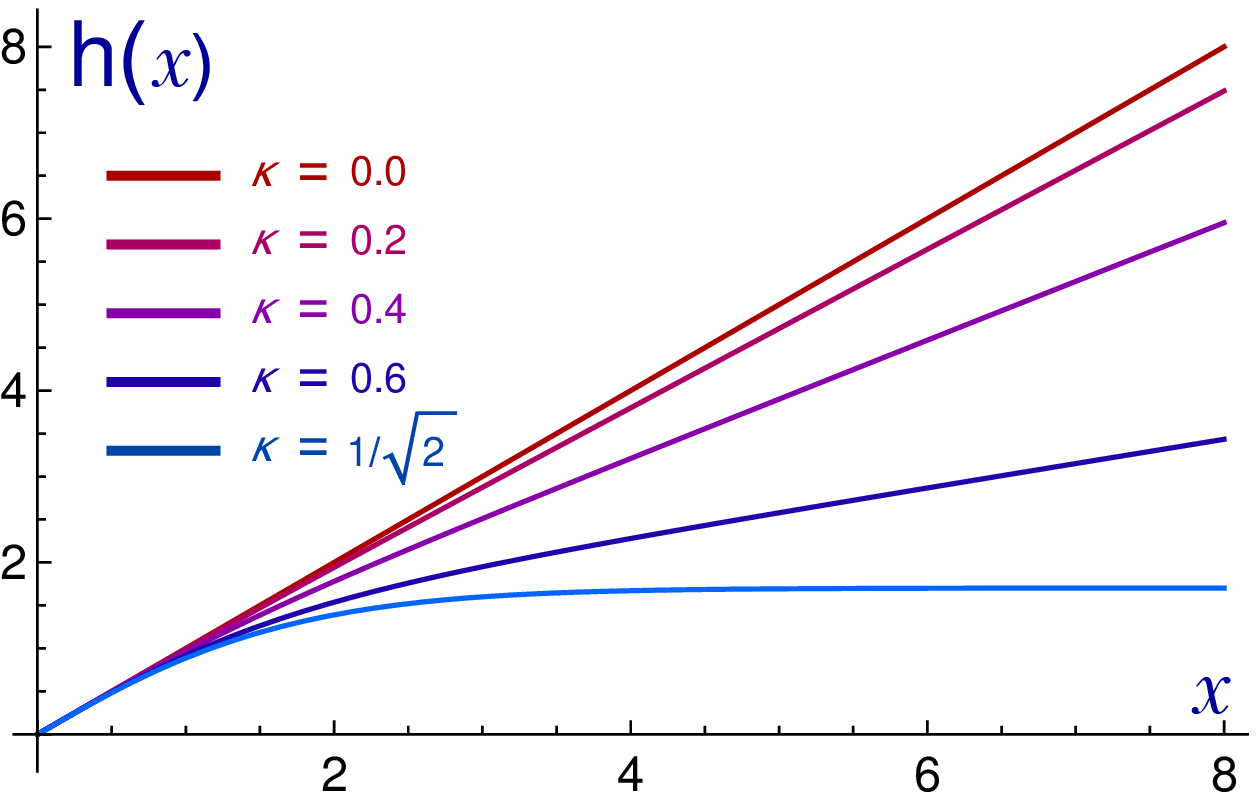}
\caption{Profiles of the energy density and the metric coefficient $h$  for $n=2$, $l=1$ and $\alpha_2=0.4$ and several strengths of the gravitational interaction.}
\label{fig6}
\end{figure}
\item Third, and last step: we develop a numerical method to find a solution interpolating between the solution close to the origin with the solution near infinity.
We integrate the self-duality equations (\ref{bograd4}) by means of an explicit Runge-Kutta method, starting from a small $x_0=10^{-6}$ value of the non-dimensional radial coordinate and seeking for the value of $\alpha_1$ in (\ref{x0}) which gives the best convergence of the fields to the boundary conditions at large $x$.
\end{enumerate}
We show some figures with results for the scalar fields $f_1(x)$ and $f_2(x)$, the magnetic field, the energy density and the metric coefficient $h(x)$. Figures \ref{fig1} and \ref{fig2} refer to the case with vorticity $n=1$ and a value of $\mathcal{H}_0=0.4$, taking for the gravitational coupling five values between $\kappa=0$ and the critical value $\kappa_{\rm crit}=1$ at the frontier of supermassive behavior. In figures \ref{fig3} and \ref{fig4}, we consider vorticity $n=2$ with $l=0$ and the same value of $\mathcal{H}_0$ than before, also with five values of the gravitational coupling growing up to the critical value $\kappa_{\rm crit}=\frac{1}{\sqrt{2}}$ corresponding to that vorticity. Finally, figures \ref{fig5} and \ref{fig6} present also results for the $n=2$ case, but now with $l=1$ and $\alpha_2=0.4$ in (\ref{x0}). As we can see from the figures, for each $x$ the scalar fields are greater for smaller gravitational coupling, while the magnetic field shows the opposite behavior. The field $f_1$ grows monotonically with $x$, but it has non-zero slope at the center of the string only for vorticity $n=1$, being the initial growing softer for higher $n$ as a consequence of (\ref{x0}). For both vorticities, $f_2$ decreases monotonically if $l=0$, but when $n=2$ and $l=1$, $f_2$ vanishes at the center of the string and grows for small values of $x$, reaching a maximum for a range of values of $x$ comparable to the range in which $f_1$ begins to be close to its vacuum expectation value. The magnetic field is maximum at $x=0$ for $n=1$ and $n=2$ with $l=1$, but in the case $n=2$ with $l=0$, the highest value of the magnetic field is displaced from the origin and configures a narrow annulus around the string center. In all cases, the energy density is zero at $x=0$ and grows up to a maximum, which is higher for lower gravitational coupling; for $n=1$ and $n=2,l=0$, the maximum is appreciably closer to the center for smaller $\kappa$, but when $n=2,l=1$, the distance from the maximum to the center is very similar for all values of $\kappa$. Also, in all cases, the metric coefficient $h(x)$ grows with $x$, reaching a clearly visible linear behavior quite soon, even before than the fields reach their asymptotic values. The slope of $h(x)$ decreases, and thus the deficit angle increases with $\kappa$, being 1 for $\kappa=0$ and 0 for $\kappa=\kappa_{\rm crit}$. 

Although all the physical properties of the self-dual cosmic string solutions in the reference chart are precisely described in the formulas and figures in this subsection we emphasize two qualitative features that distinguish them as belonging to a first species of string defects in this system:
\begin{enumerate}

\item All the energy lumps associated to these solutions are concentrated in the spatial plane perpendicular to the string, either decaying to their vacuum values exponentially when $f_2(\rho)=0$ or more spreaded throughout the spatial plane if $f_2(\rho)\neq 0$. The crux of the matter is that all of them are mapped into the subset of the reference chart of $\mathbb{CP}^2$ comprised between the point ($f_1(0)=0$, $f_2(0)=\delta_{l0}\mathcal{H}_0$) and the $\mathbb{S}^3$-vacuum orbit.

\item In the case $f_2(x)=0$ the situation is easier to visualize because the system reduces to target in a $\mathbb{CP}^1$ submanifold of $\mathbb{CP}^2$, the vacuum orbit becomes the paralel $\vert \psi_1^{(v)}\vert^2=v^2$ and the solution is mapped into the spherical cap bounded by this parallel and the South Pole $\vert \psi_1\vert=0$.

\item Any attempt to prolongate these cosmic strings to the North hemisphere would demand infinite energy because the fields should trespass the vacuum orbit. 

\end{enumerate}

$\mathbb{CP}^2$ instantons appear in our model when we decouple gravity ($G=0$) and the gauge field ($e=0$). They  are similar to the solutions found in this subsection in that in both cases there is a modulus ruling their spatial size, the $\omega$ parameter of (\ref{instanton}) for instantons and ${\cal H}_0$ for cosmic strings, although the origin of the instanton modulus is scale invariance and this symmetry is lost for cosmic strings. Objects of both types show however appreciable differences in the range that they cover on target space: the scalar fields of cosmic strings take values on a region of $\mathbb{CP}^2$ bounded by the vacuum orbit, whilst instantons spread over all $\mathbb{CP}^2$. As commented before, this makes instantons infinitely massive when we take $e>0$ to obtain cosmic strings.
\subsection{Multi-center self-dual cosmic strings}
A generalization of the Atiyah-Singer index theorem led E. Weinberg in \cite{Weinberg} to show that the moduli space of self-dual vortices in the AHM has dimensions $2n$ were $n$ is the vorticity. This means that there is freedom of placing the centers of the flux tubes with a quantum of magnetic flux throughout the plane. The explicit zero modes corresponding to this freedom were found and explicitly constructed in Reference \cite{AAlonso} for the AHM. Identical task was performed in Reference \cite{AAlonso1} about the vortex zero modes in the non-linear $\mathbb{CP}^1$-sigma model. In both papers we started from the cylindrically symmetric multi-vortex solutions to find their deformations respecting self-duality. The index theorem showing that the dimension of the moduli space of self-dual vortices in the semi-local AHM is $4n$ was developed in References \cite{AAlonso2}-\cite{AAlonso3}.
The construction of the general self-dual vortices in the semi-local AHM in terms of the $n$-centers of the magnetic flux tubes plus other $n$ complex parameters characterizing the behaviour of the $\psi_2$ field was achieved in Reference \cite{giorrusa92}. It is natural to explore how much this scheme survives the coupling to gravity.  Thus, let us now briefly consider the existence of self-dual solutions beyond those with  cylindrical symmetry. For this purpose, it is convenient to adopt for the metric on the transverse plane the conformal gauge $\gamma_{ij}=e^{-4V(x)}\delta_{ij}$ \cite{lin88}. This gauge is advantageous in that the Bogomolny equation (\ref{bog2}) takes in it the same form $D_1\psi_p+iD_2\psi_p=0$ than in Euclidean coordinates, and thus we can use the Poincar\'e $\bar{\partial}$-lemma to guarantee that the solution of (\ref{bog2}) for $p=1$ has exactly $n$ zeros on the plane, counted with multiplicity \cite{jatau80}.  Let us denote these zeros by $z_1,z_2,\ldots,z_n$, where we are now using a complex coordinate $z=x^1+i x^2$. Also, (\ref{bog2}) for $p=2$  implies that $\psi_2=w(z)\psi_1$, where \cite{giorrusa92}
\bdm
w(z)=\frac{Q_n(z)}{P_n(z)},\hspace{1cm} P_n(z)=\prod_{s=1}^n(z-z_s),\hspace{1cm} Q_n(z)=q_{n-1} z^{n-1}+\cdots+q_1 z+q_0.
\edm
Using these expressions, the remaining Bogomolny equation (\ref{bog1}) may be rewritten in the form
\beq
\Delta\log|\psi_1|=e^2 v^2 e^{-4V}\frac{|\psi_1|^2(1+|w|^2)-v^2}{|\psi_1|^2(1+|w|^2)+v^2}+2\pi\sum_{s=1}^n\delta^{(2)}(z-z_s),\label{boglap}
\eeq
where $\Delta$ denotes here the Euclidean Laplacian, while the Einstein equation (\ref{einsteinautodual}) is
\bdm
\Delta\Big[ V-4\pi Gv^2\Big(\log{\cal D}-\frac{1}{2}\log\frac{|\psi_1|^2}{\prod_{s=1}^n|z-z_s|^2}\Big)\Big]=0.
\edm
This means that the expression into the brackets is harmonic and bounded, and hence a constant. Thus, we obtain for the conformal factor exponent $V(x^1, x^2)$
\beq
V=4\pi Gv^2\Big[\log\left(v^2+|\psi_1|^2(1+|w|^2)\right)-\frac{1}{2}\log\frac{|\psi_1|^2}{\prod_{s=1}^n|z-z_s|^2}{\Big]}+{\rm constant}.\label{soluV}
\eeq
Therefore, a solution of (\ref{boglap}) with $V(x^1,x^2)$ given by (\ref{soluV}) and such that $\psi_1\rightarrow v$ for $|z|\rightarrow\infty$ and $\psi_1=0$ at the $n$ zeros $z_s$ represents a system of separated and parallel cosmic strings which are centered around the positions $z=z_s$. Of course, to show rigorously that (\ref{boglap}) with these boundary conditions has an unique solution is a difficult problem in functional analysis which is beyond the scope of this work, but we can nevertheless argue through the heuristic physical arguments pointed in \cite{lin88} that this is the case. We saw in previous subsection that out of the core of the vortex the metric converges very quickly to the Minkowskian form with a deficit angle, meaning that  well separated strings at rest do not feel any mutual gravitational forces. Therefore, the existence of a state of equilibrium for such a system depends only on the balance between the $U(1)$ and scalar interactions among the strings, and given that gravity does not enter in this balance, we can study it in the limit in which the gravitational coupling $G$ vanishes. $G=0$ implies $V=0$ and with the change of variables
\bdm
\xi=\log\left(|f_1|^2(1+|w|^2)\right)-u_1,\hspace{1cm}u_1=\log\left(|P_n(z)|^2+|Q_n(z)|^2\right)-\sum_{s=1}^n\log\left(\nu+|z-z_s|^2\right)
\edm
where $\nu$ is an arbitrary constant, equation (\ref{boglap}) becomes
\beq
\Delta \xi=g_0+ \tanh\frac{u_1+\xi}{2}\hspace{2cm} g_0=4\sum_{s=1}^n\frac{\nu}{\nu+|z-z_s|^2}.\label{ecuacion para v}
\eeq
This is the variational equation of the functional
\beq
a(\xi)=\int d^2z\left\{\frac{1}{2}|\nabla \xi|^2+g_0 \xi+2 \log\left(\cosh\frac{u_1+\xi}{2}\right)\right\}\label{funcional}
\eeq
and the proof that (\ref{ecuacion para v}) has an unique solution with $\xi\rightarrow 0$ for $|z|\rightarrow\infty$ and $\xi\rightarrow -\infty$ at the $n$ zeros $z_s$ proceeds through the analysis of this functional along the lines developed by Taubes \cite{jatau80} for the AHM. Notice, however, that for $|z|\rightarrow\infty$ and at the points $z_s$, (\ref{funcional}) and the functional corresponding to standard semi-local vortices coincide. It is thus to be expected that the result of the analysis is the same in this case than for for the semi-local model \cite{giorrusa92}, leading also for the $\mathbb{C}P^2$ model to the existence of multivortex solutions. In fact, as explained in \cite{giorrusa92}, the crucial points of the proof are to show that $a(\xi)$ is strictly convex and that its radial derivative $(D_\xi a)(\xi)$ is positive for a sufficiently large ball in the space of $\xi$ functions. To check the first point, it is convenient to adopt the notation of \cite{jatau80} to write
\bdm
a(\xi)=\frac{1}{2}\|\nabla \xi\|^2+\langle g_0,\xi\rangle+2 \left\langle 1,\log\left(\cosh\frac{u_1+\xi}{2}\right)\right\rangle
\edm
with $L^2(\mathbb{C})$ norm and scalar product. Then, given that the first term is quadratic, hence strictly convex, the second is linear and the third has positive second derivative respect $\xi$, and it is thus also convex, it follows that $a(\xi)$ is strictly convex. With regard to the second point, the radial derivative of $a(\xi)$ is
\bdm
(D_\xi a)(\xi)=\|\nabla \xi\|^2+M(\xi)\hspace{1cm} M(\xi)=\left\langle \frac{\xi}{e^{u_1+\xi}+1}, g_0-1+e^{u_1+\xi}(1+g_0)\right\rangle
\edm
and has the same structure than in the standard semi-local case, in which
\bdm
M_{\rm SML}(\xi)=\left\langle \xi,g_0-1+e^{u_1+\xi}\right\rangle.
\edm
Now, by taking $\nu$ large enough, it is easy to see that negative contributions to $M(\xi)$ are always less important than those appearing in $M_{\rm SML}(\xi)$. Thus, like in the standard semi-local case, one can prove that the radial derivative is positive and there is an unique solution of (\ref{boglap}).
\section{Type I and Type II versus self-dual cosmic strings}
\begin{figure}[H]
\centering
\includegraphics[width=5.5cm]{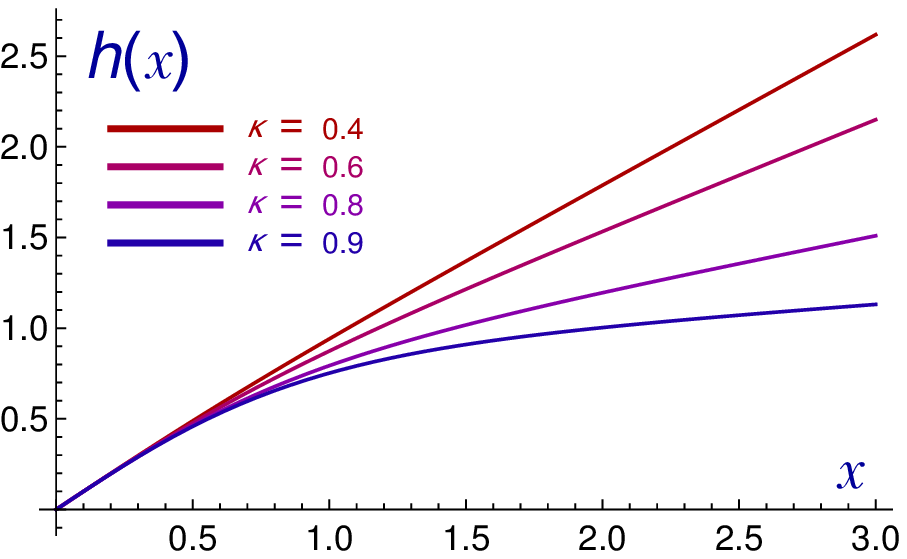}
\hspace{1.5cm}
\includegraphics[width=5.5cm]{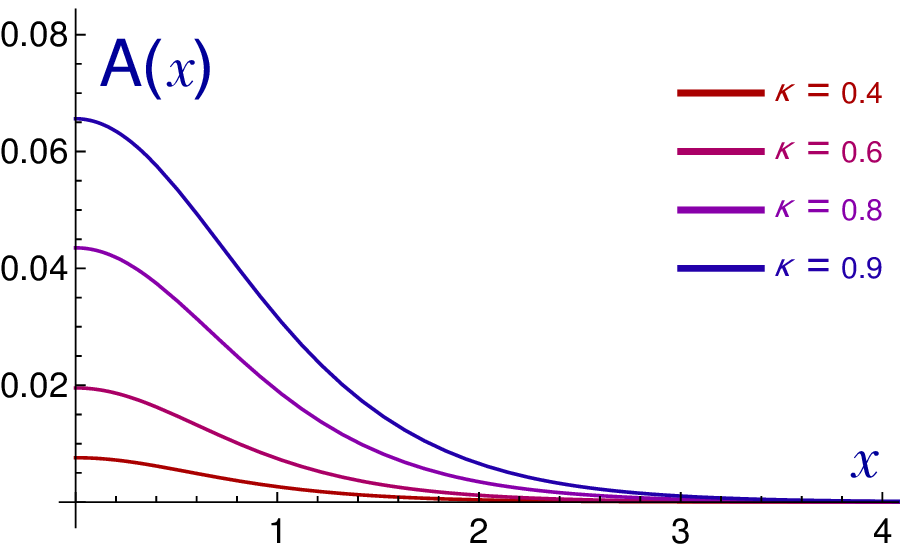}
\caption{The metric coefficients $A(x)$ and $h(x)$ for $\mathcal{H}_0=0$, $\beta=1.5$ and several strengths of the gravitational interaction.}
\label{fig7}
\end{figure}

\begin{figure}[H]
\centering
\includegraphics[width=5.5cm]{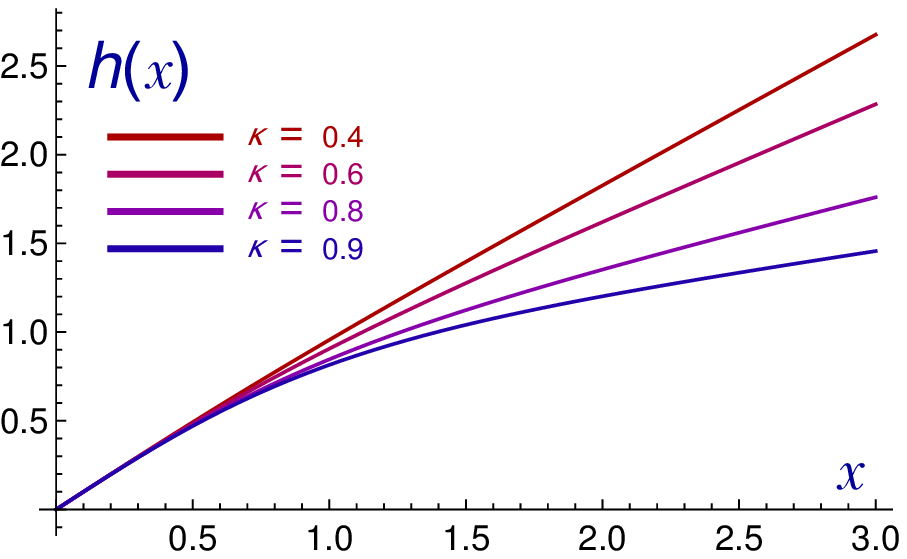}
\hspace{1.5cm}
\includegraphics[width=5.5cm]{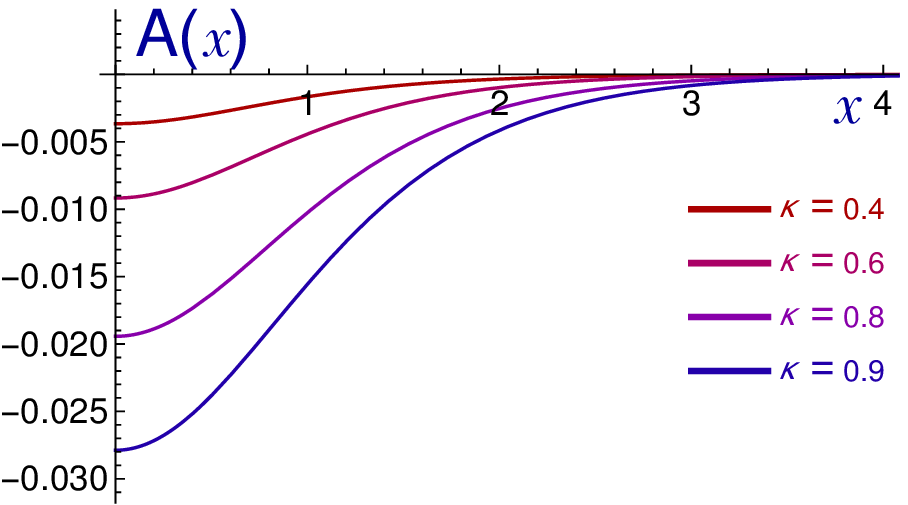}
\caption{The metric coefficients $A(x)$ and $h(x)$ for $\mathcal{H}_0=0$, $\beta=0.8$ and several strengths of the gravitational interaction.}
\label{fig8}
\end{figure}

We come back in this Section to the second-order ODE system obtained at the end of Section 2, taking,  as in the Section 3, ${\cal E}(|\psi|)=1$ for the dielectric function but now the scalar potential is non critically coupled, i.e.,
\beq
U(|\psi|)\equiv\frac{1}{2} W^2(|\psi|)=\frac{\lambda v^4}{2}\left(\frac{|\psi_1|^2+|\psi_2|^2-v^2}{|\psi_1|^2+|\psi_2|^2+v^2}\right)^2 \label{potlambda}
\eeq
with $\lambda\neq e^2$. Substitution of the explicit form of the Fubini-Study metric in the Euler-Lagrange equations (\ref{eqelrad1})-(\ref{eqelrad2}) brings about the cancellation of all phases and then, passing to non-dimensional variables, the equations become
\beq
\frac{d}{dx}\left(\frac{e^A}{h}\frac{da}{dx}\right)= 2\frac{e^A}{h}\left[\frac{2}{\mathfrak D}\sum_{p=1}^2 {\cal A}_p f_p^2-\frac{1}{{\mathfrak D}^2}\\\sum_{p=1}^2\sum_{l=1}^2\left({\cal A}_p+{\cal A}_l\right)f_p^2 f_l^2\right]\label{elrad1}
\eeq
\beqr
\frac{1}{h e^A}\frac{d}{dx}\left\{h e^A\left[\frac{1}{\mathfrak D}\frac{d f_q}{dx}-\frac{f_q}{{\mathfrak D}^2}\sum_{l=1}^2 f_l \frac{df_l}{dx}\right]\right\}-\frac{{\cal A}_q^2 f_q}{h^2 {\mathfrak D}}+\frac{{\cal A}_q f_q}{h^2 {\mathfrak D}^2}\sum_{l=1}^2 {\cal A}_l f_l^2=\nonumber\\-\frac{1}{{\mathfrak D}^2}\left[\sum_{l=1}^2{\cal X}_{ll} f_q+\sum_{l=1}^2{\cal X}_{ql} f_l-\frac{2}{\mathfrak D}\sum_{p=1}^2\sum_{l=1}^2{\cal X}_{pl} f_p f_l f_q\right]+\beta \frac{{\mathfrak D}-2}{{\mathfrak D}^3} f_q,\label{elrad2}
\eeqr
where there is no sum on repeated $q$ indices, $\beta=\frac{\lambda}{e^2}$ and
\bdm
{\mathfrak D}=1+ f_1^2+f_2^2\hspace{2cm} {\cal X}_{pl}=\frac{df_p}{dx}\frac{df_l}{dx}+\frac{{\cal A}_p {\cal A}_l}{h^2} f_p f_l\quad({\rm no\ sum}).
\edm
Also in non-dimensional variables, the Einstein equations (\ref{einsteint}) and (\ref{einsteinphi}) are
\beqr
\frac{1}{h}\Big[\frac{d^2h}{dx^2}+\frac{1}{2}\frac{d}{dx}\Big(h\frac{da}{dx}\Big)\Big]+\frac{1}{4}\Big(\frac{dA}{dx}\Big)^2&=&-\frac{\kappa^2}{2}(\varepsilon_a+\varepsilon_f+\varepsilon_{af}+u)\label{einsteinrad1}\\
\frac{d^2A}{dx^2}+\frac{3}{4}\left(\frac{d A}{dx}\right)^2&=&\frac{\kappa^2}{2}(\varepsilon_a-\varepsilon_f+\varepsilon_{af}-u)\label{einsteinrad2}
\eeqr
where the different contributions to the energy-momentum tensor turn out to be
\beqrn
\varepsilon_a&=&\frac{1}{h^2}\left(\frac{da}{dx}\right)^2\hspace{5.3cm}\varepsilon_f=\frac{4}{{\mathfrak D}}\left[\sum_{l=1}^2\left(\frac{df_l}{dx}\right)^2-\frac{1}{\mathfrak D}\left(\sum_{l=1}^2 f_l\frac{df_l}{dx}\right)^2\right]\\
\varepsilon_{af}&=&\frac{4}{h^2{\mathfrak D}}\left[\sum_{l=1}^2{\cal A}_l^2 f_l^2-\frac{1}{\mathfrak D}\left( \sum_{l=1}^2 {\cal A}_l f_l^2\right)^2\right]\hspace{0.8cm}u=\beta\left(\frac{{\mathfrak D}-2}{\mathfrak D}\right)^2.
\eeqrn
Thus, to find non self-dual cosmic strings we have to solve the system (\ref{elrad1}), (\ref{elrad2}), (\ref{einsteinrad1}) and (\ref{einsteinrad2}) with boundary conditions (\ref{contorno1}), (\ref{contorno2}), (\ref{contorno3}) and (\ref{contorno4}), but given that the equations are invariant under the shift of $A(x)$ by a constant, it is more convenient for the shooting method to change (\ref{contorno3}) by the initial condition
\beq
A(0)=0\hspace{2cm} A^\prime(0)=0,\label{recontorno3}
\eeq
since for a solution approaching Minkowski space with a deficit angle at large distances, Einstein equations ensure that $A(x)$ will be constant at infinity and thus, subtracting this constant from the solution obtained with (\ref{recontorno3}), we recover the behavior (\ref{contorno3}). We will concentrate on the case with $l=0$, so that from the boundary conditions at the origin we can assume that for very small $x$ the fields are of the form
\beqrn
f_1(x)&\simeq&\alpha_1 x^{r_1}\hspace{2cm} f_2(x)\simeq\mathcal{H}_0+\alpha_2 x^{r_2}\hspace{2cm} a(x)\simeq n-\delta x^s\\
h(x)&\simeq&x\hspace{2.7cm}A(x)\simeq\eta x^t
\eeqrn
with positive exponents and $t\geq 2$. The difference with the self-dual case is that now the equations are second-order and  we need to fix not only the field values, but also their derivatives, for some $x_0$ very close to zero, so that although $l=0$ we have to retain the second term in $f_2(x)$. Substituting these expressions in (\ref{elrad1})-(\ref{einsteinrad2}) and working to leading order, we find $r_1=n$ and $r_2=s=2$ like in the self-dual situation, and also
\bdm
t=2\hspace{1.5cm}\alpha_2=-\frac{1}{4}\frac{1-\mathcal{H}_0^2}{1+\mathcal{H}_0^2} \mathcal{H}_0\beta\hspace{1.5cm} \eta=\kappa^2\left[\delta^2-\frac{1}{4}\left(\frac{1-\mathcal{H}_0^2}{1+\mathcal{H}_0^2}\right)^2 \beta\right].
\edm
We have thus two free parameters, $\alpha_1$ and $\delta$, and we shall proceed by a shooting procedure in which we will take $x_0=10^{-6}$ and will try to adjust the values of these parameters in such a way that for large $x$ the fields behave consistently with the boundary conditions at infinity.
In particular, an interesting case to be considered is that of $\mathcal{H}_0=0$. In that situation the target manifold of the scalar fields reduces to $\mathbb{CP}^1$, or $\mathbb{S}^2$, and we have thus cosmic strings with values on the sphere. Vortices and cosmic strings on the sphere were previously studied by means of the Bogomolny equations in \cite{nosotros15,nivi12,vemadlar03} and we will here revisit this problem to provide some new results for the non self-dual case, where the main novelty is the presence of a non-vanishing function $A(x)$. Thus, we show in figures \ref{fig7} and \ref{fig8} the profile of this function, and also of the other metric coefficient $h(x)$, where we have considered the effect of several values of the gravitational coupling for scalar potentials of type I and II, namely with $\beta=0.8$ and $\beta=1.5$. As one can see from the figures, the space-time of the cosmic string becomes Minkowskian at distances quite close to the string center and, for fixed gravitational coupling, the slope of $h(x)$ becomes smaller, and thus the deficit angle greater, as $\beta$ increases. As for $A(x)$, we see that for values of the coupling $\beta$ in the range studied, the departure of this function with respect to the self-dual regime $A(x)=0$ is rather small, and that $A(x)$ decreases monotonically from positive values to zero when $\beta>1$, whereas its performance is just the opposite for  $\beta<1$.

\begin{figure}[t]
	\centering
	\includegraphics[height=3.cm]{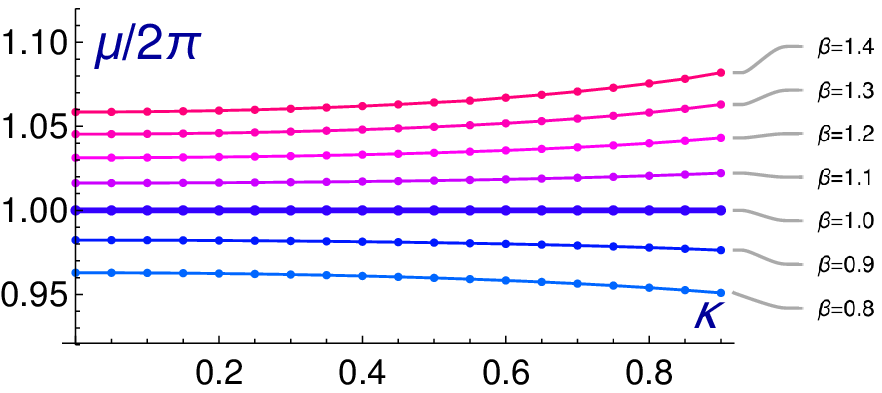}
	\hspace{0.8cm}
	\includegraphics[height=3.cm]{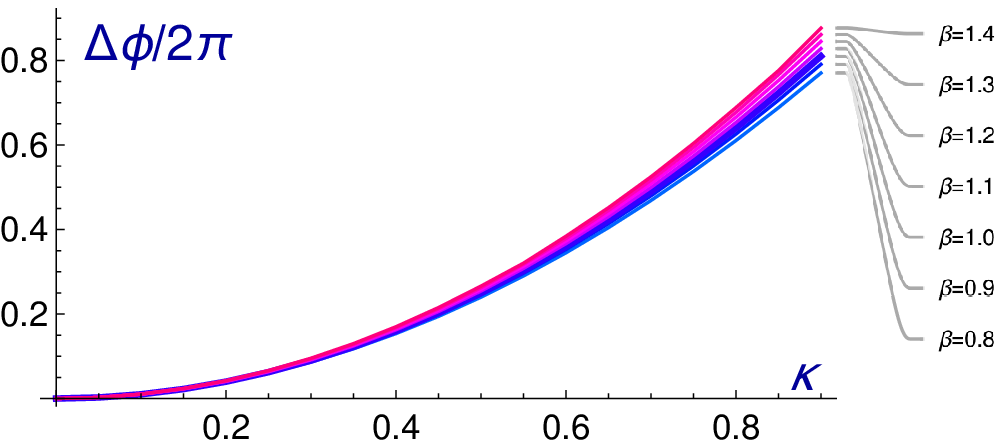}
	\caption{Energy per unit length (left) and deficit angle (right) normalized to $2\pi$ as a function of $\kappa$ and several values of $\beta$ for the case $\mathcal{H}_0=0$ .}
	\label{figtabla1}
\end{figure}

We have also computed the values of the energy per unit length  and the deficit angle for cosmic strings on the sphere in several cases, and we show the results in Tables \ref{table1}  and \ref{table2}. As we can see, for each fixed value of the coupling $\kappa$, $\mu$ and $\Delta\phi$ increase with $\beta$, a behavior that in the second case was also clearly observed in the Figures \ref{fig7} and \ref{fig8}. Moreover, while the deficit angle always increases as the gravitational coupling grows, the energy per unit length increases or decreases with $\kappa$ depending on whether $\beta>1$ or $\beta<1$.
Finally, in Table \ref{table3} we present the values of the energy per unit length of non self-dual vortices on the sphere with $\kappa=0$ for several vorticities. We observe that for $\beta>1$ the energy of a vortex with vorticity $n$ is higher than $n$ times the energy of the vortex of unit vorticity, while in the case $\beta<1$ we obtain the opposite behavior, see Figure \ref{figtabla1}. This confirms the character of superconductivity in this system in the usual way: for $\beta>1$ the forces between vortices are repulsive and the superconductivity is of type II, while forces are attractive for $\beta<1$ giving rise to Type-I superconductivity.
\begin{table}
\begin{center}
\begin{tabular}{||l||c|c|c|c|c|c||}
\hline
\multicolumn{7}{||c||}{Energy per unit length $(\frac{\mu}{2\pi})$}\\
\hline
$\kappa$&0&0.2&0.4&0.6&0.8&0.9\\
\hline
$\beta=0.8$&0.963&0.962&0.961&0.958&0.954&0.951\\
\hline
$\beta=1$&1.000&1.000&1.000&1.000&1.000&1.000\\
\hline
$\beta=1.5$&1.071&1.072&1.075&1.081&1.092&1.100\\
\hline
\end{tabular}
\end{center}
\caption{Energy per unit length normalized to $2\pi$ for $\mathcal{H}_0=0$ and several values of $\beta$ and $\kappa$.}
\label{table1}
\end{table}
\begin{table}
\begin{center}
\begin{tabular}{||l||c|c|c|c|c|c||}
\hline
\multicolumn{7}{||c||}{Deficit angle $(\frac{\Delta\phi}{2\pi})$}\\
\hline
$\kappa$&0&0.2&0.4&0.6&0.8&0.9\\
\hline
$\beta=0.8$&0.000&0.038&0.154&0.345&0.611&0.770\\
\hline
$\beta=1$&0.000&0.040&0.160&0.360&0.640&0.810\\
\hline
$\beta=1.5$&0.000&0.043&0.172&0.389&0.699&0.891\\
\hline
\end{tabular}
\end{center}
\caption{Deficit angle normalized to $2\pi$ for $\mathcal{H}_0=0$ and several values of $\beta$ and $\kappa$.}
\label{table2}
\end{table}
\begin{table}[b]
\begin{center}
\begin{tabular}{||l||c|c|c|c||}
\hline
\multicolumn{5}{||c||}{Energy per unit length $(\frac{\mu}{2\pi})$}\\
\hline
$n$&1&2&3&4\\
\hline
$\beta=0.8$&0.963&1.897&2.821&3.740\\
\hline
$\beta=1$&1.000&2.000&3.000&4.000\\
\hline
$\beta=1.5$&1.071&2.203&3.357&4.522\\
\hline
\end{tabular}
\end{center}
\caption{Energy per unit length normalized to $2\pi$ for vortices on the sphere of several vorticities without gravitational coupling}
\label{table3}
\end{table}

\begin{figure}[t]
\centering
\includegraphics[height=3.cm]{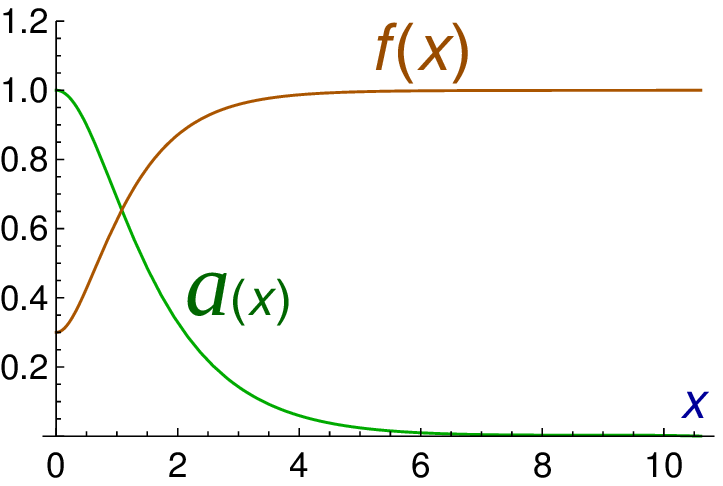}
\hspace{0.2cm}
\includegraphics[height=3.cm]{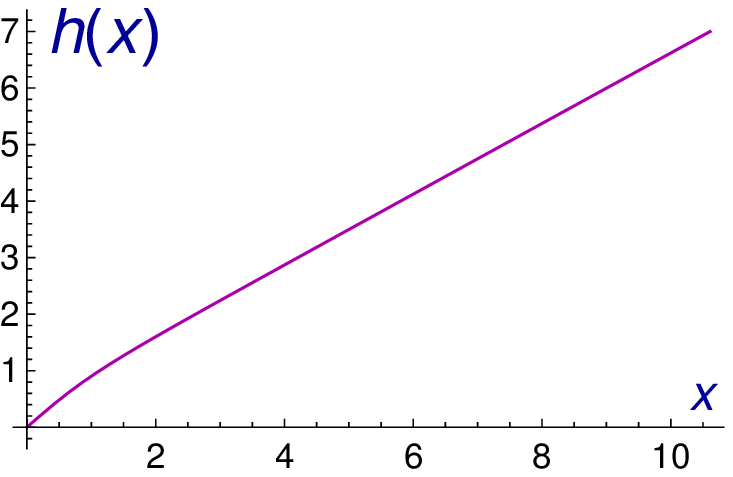}
\hspace{0.2cm}
\includegraphics[height=3.cm]{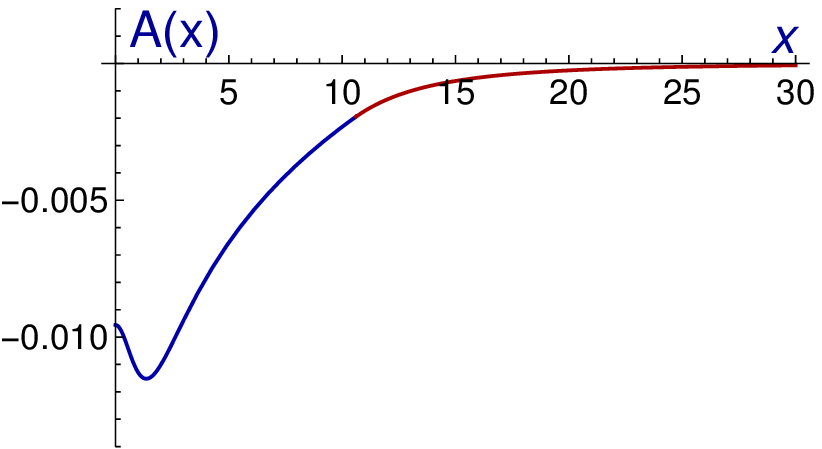}
\caption{Profiles of $a(x)$, $f(x)$, $h(x)$ and $A(x)$ for the case $n=1$, $\mathcal{H}_0=0.3$, $\beta=1.2$ and $\kappa=0.6$.}
\label{fig9}
\end{figure}

\begin{figure}[H]
	\centering
	\includegraphics[height=3.cm]{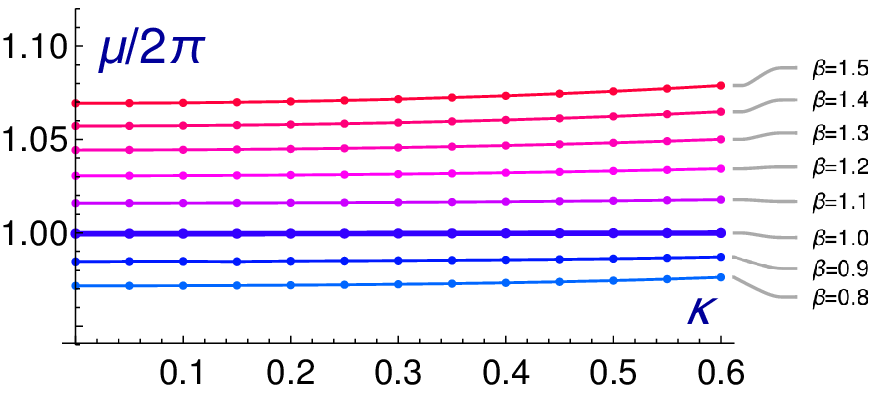}
	\hspace{0.8cm}
	\includegraphics[height=3.cm]{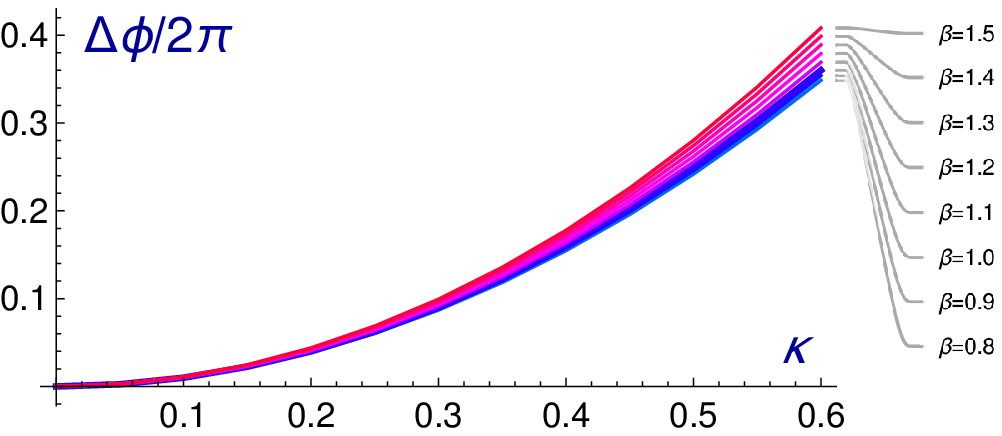}
	\caption{Energy per unit length (left) and deficit angle (right) normalized to $2\pi$ as a function of $\kappa$ for several values of $\beta$ with ${\cal H}_0=0.3$.}
	\label{figtabla2}
\end{figure}

When $\mathcal{H}_0\neq 0$ it is more difficult to make the shooting converge properly at large distances, because in these cases the fields approach their asymptotic values as inverse powers of $x$ instead of exponentially. Typically, by fixing the parameters $\alpha_1$ and $\delta$ with a precision of nine decimal digits, one obtains a solution which behaves correctly for $x$ up to around 10 or 12, but to extend the solution beyond that range requires much more precision and computer time. In general, the rates of convergence of the scalar and gauge fields, and that of the metric tensor coefficient $h(x)$, are quite satisfactory, while $A(x)$ meets the expected behavior at infinity much poorly. To palliate this problem, the pure shooting method has to be supplemented by an estimation of the fields at large distances. In fact, as one can check using equations (\ref{elrad1}), (\ref{elrad2}), for $\beta\neq 1$ the dominant contributions to the fields at high $x$ have the same form (\ref{infinito}) than in the self-dual case, and bearing this in mind we can approximate the component $T^\phi_{\;\phi}$ of the energy-momentum tensor for large $x$ as
\bdm
 T^\phi_{\;\phi}\simeq -\frac{1}{2}\frac{C^2 n^2}{r^2} x^{-2(\frac{n}{r}+1)}
\edm
at leading order. Using this result in equation (\ref{einsteinrad2}), we obtain for $A(x)$, also at leading order and far from the origin,
\beq
A(x)\simeq-\frac{\kappa^2 C^2 n^2}{4n(2n+r)}x^{-\frac{2n}{r}}+{\rm constant},\label{suplemento}
\eeq
an expression that can be employed to extend the results obtained by shooting to longer $x$ when necessary. As an example, we show in Figure \ref{fig9} the results of the shooting method with $\alpha_1$ and $\delta$ adjusted to nine decimal figures for the case $n=1$, $\mathcal{H}_0=0.3$, $\beta=1.2$ and $\kappa=0.6$. As we can see from the figure, the convergence of $f(x)=\sqrt{f_1^2(x)+f_2^2(x)}$, $a(x)$ and $h(x)$ is quite convincing, but $A(x)$ does not attains so quickly its constant asymptotic value: the blue part of the profile, obtained by the shooting, shows still a slope of 0.00059,  for an absolute value of 0.0076, even when the fields have already reached very close values to their vevs.  This can be repaired by means of (\ref{suplemento}), which has been used to draw the red part of the graphic. Thus, while for small values of $\mathcal{H}_0$ we expect a form of $A(x)$ similar to those found for cosmic strings on the sphere, when $\mathcal{H}_0$ is higher the pattern is different: in the case at hand, $A(x)$ begins decreasing near $x=0$, but it arrives to a minimum for finite $x$ and then increases asymptotically towards zero, keeping constant negative sign for all $x$.
\begin{table}[t]
\begin{center}
\begin{tabular}{||l||c|c|c|c|c|c||}
\hline
\multicolumn{7}{||c||}{Energy per unit length $\left(\frac{\mu}{2\pi}\right)$}\\
\hline
$\kappa$&0.0&0.2&0.3&0.4&0.5&0.6\\
\hline
$\beta=0.8$&0.974&0.974&0.975&0.977&0.980&0.984\\
\hline
$\beta=1$&1.000&1.000&1.000&1.000&1.000&1.000\\
\hline
$\beta=1.2$&1.037&1.037&1.040&1.042&1.047&1.055\\
\hline
\end{tabular}
\end{center}
\caption{Energy per unit length normalized to $2\pi$ for several values of $\beta$ and $\kappa$ for the case with $\mathcal{H}_0=0.3$}
\label{table4}
\end{table}
% \begin{figure}[H]
% \centering
% \includegraphics[width=5.5cm]{figura10.eps}
% \caption{Profile of $A(x)$  for the case $n=1$, $h_0=0.3$,$\beta=1.2$ and $\kappa=0.6$.}
% \label{fig10}
% \end{figure}
As some further examples of the behavior of the system for $\mathcal{H}_0\neq 0$, we present several results for the case $\mathcal{H}_0=0.3$ in Tables \ref{table4} and \ref{table5}. To obtain the values shown in Table \ref{table4}, we have taken into account that (\ref{infinito}) implies that $ T^t_{\;t}(x)\propto x^{-\frac{2n}{r}-2}$ at leading order for large $x$. Thus, if we define $\hat{\mu}(y)=-\int_0^y dx h(x) T^t_{\;t}(x)$, we can estimate
\bdm
\frac{\mu}{2\pi}=\hat{\mu}(y)+\frac{1}{2} r \hat{\mu}^\prime(y) y,
\edm
where for $y$ we substitute the  larger $x$ value for which the shooting method achieves a good convergence, and $\hat{\mu}^\prime(y)$ is the corresponding value of the integrand. In the case of Table \ref{table5}, we use the known behaviors of $T^t_{\;t}(x)$ and $A(x)$ for large $x$ and equation (\ref{einsteinrad1}) to approach $h^\prime(x)\simeq r-\left(\frac{2}{r}-1\right) b x$, with some constant $b$, in that zone, and from this it follows the estimate
\bdm
\frac{\Delta\phi}{2\pi}=1-\frac{2 h^\prime(y)}{2-h^{\prime\prime}(y) y}
\edm
with the same $y$ than before. At any rate, we find a similar pattern than that described previously, see Figure \ref{figtabla2}. 

Let us finally comment on the issue of stability. Viewed as solutions on the sphere, cosmic strings with $\mathcal{H}_0=0$ are stable by topological reasons, although for type II superconductivity  those  with $n>1$ tend to break up into $n=1$ components due to repulsive forces. Their stability status, however, is different when they are embedded in the full $\mathbb{CP}^2$ model. For the standard semi-local model without gravity, the analysis of the stability of Nielsen-Olesen vortices (i.e. with $\psi_2=0$) of topological number $n=1$ was carried out in \cite{hind92}, and that paper offers also an explanation of the results based on the following heuristic argument. When we perturbe a Nielsen-Olesen vortex adding a non-null $\psi_2$ component, new terms in the energy per unit length appear; those coming from the covariant derivatives are always positive, giving rise to an increase in energy, but the presence of a $\psi_2\neq 0$ component makes the potential energy 
\bdm
U(|\psi|)=\frac{\lambda v^4}{2}\left(|\psi_1|^2+|\psi_2|^2-v^2\right)^2
\edm
to go down. If the coupling $\lambda$ in the potential is great enough the effect of the potential prevails, the energy per unit length decreases and the Nielsen-Olesen solution turns out to be unstable. On the other hand, when $\lambda\rightarrow 0$ the potential is very small and the most important contribution comes from gradient energy, making the Nielsen-Olesen solution stable. The boundary between both regimes is precisely the self-dual limit $\lambda=e^2$, because in this case we know from the Bogomolny analysis that $\mathcal{H}_0$ is a modulus and the energy is the same for $\psi_2$ null or different from zero: all vortices are in neutral equilibrium by saturating the lower bound $\mu=2\pi v^2$. In the $\mathbb{CP}^2$ model the terms coming from covariant derivatives are more complicated, but the behavior of the potential energy (\ref{potlambda}) with regards to the addition of a $\psi_2$ component to the Nielsen-Olesen solution is exactly the same than in the standard semi-local case. Therefore, like there, the $\mathbb{CP}^2$ solutions with $\mathcal{H}_0=0$ and high $\beta$ have an unstable mode towards developing a second scalar component, with the accompanying dispersion of the magnetic flux. Also like there, this behavior has to cease when $\beta=1$ due to the saturation of the Bogomolny bound. Thus, we expect that the self-dual limit of the $\mathbb{CP}^2$ theory is again the borderline between unstable and stable Nielsen-Olesen solutions for, respectively, type II and type I superconductivity. This reasoning does not depend on the presence or absence of gravity and, in any case, the qualitative behavior without gravity backreaction should still be valid for small gravitational coupling. Although a complete justification of these expectations requires a detailed analysis of the Hessian spectrum, the results shown in Figures \ref{figtabla1} and \ref{figtabla2} seem to point in the correct direction. For instance, we can see in these figures that the type I solution with $\beta=0.8$ and $\kappa=0.6$ displays a lower energy per unit length when $\mathcal{H}_0=0$ than for $\mathcal{H}_0=0.3$, while in the case of the type II solution with $\beta=1.4$ and the same $\kappa$ the situation is the opposite. 

At any event, it is stressed in \cite{hind92} that the instability of type II Nielsen-Olesen vortices does not preclude the existence of some metastable semi-local vortex solution with $\beta>1$ in which the magnetic flux disperses throughout some broad but finite core. From this perspective, detailed study of type II solutions with $\mathcal{H}_0\neq 0$ is interesting.
\begin{table}[t]
\begin{center}
\begin{tabular}{||l||c|c|c|c|c|c||}
\hline
\multicolumn{7}{||c||}{Deficit angle $\left(\frac{\Delta\phi}{2\pi}\right)$}\\
\hline
$\kappa$&0.0&0.2&0.3&0.4&0.5&0.6\\
\hline
$\beta=0.8$&0.000&0.039&0.087&0.156&0.243&0.352\\
\hline
$\beta=1$&0.000&0.040&0.090&0.160&0.250&0.360\\
\hline
$\beta=1.2$&0.000&0.042&0.094&0.167&0.263&0.382\\
\hline
\end{tabular}
\end{center}
\caption{Deficit angle  normalized to $2\pi$ for several values of $\beta$ and $\kappa$  for the case with $\mathcal{H}_0=0.3$}
\label{table5}
\end{table}
\section{Changing charts: a second species of semi-local $\mathbb{CP}^2$ cosmic strings}

In this Section we investigate cosmic strings living in a second chart of the minimum atlas of $\mathbb{CP}^2$: the structure of the cosmic string manifold in the third chart is identical to the structure in the second chart,\,  i.e., cosmic strings of the second species live both in the second and third chart. We choose the transition functions from the previous chart $V_\psi$ to the new chart $V_\phi$
in their intersection $V_\psi \cap V_\phi$ as follows:
\begin{equation}
\phi_1=\frac{v^2}{\psi_1^*} \quad , \quad \phi_2=v\frac{\psi_2^*}{\psi_1^*} \, .
\end{equation}
Note that we use in the definition of the transition function the conjugates of the original fields. The reason is that we want the quanta to carry the same electric charge in all the charts. The K$\ddot{\rm a}$hler potential looks slightly different
\[
K=4 v^2\left[\log\left(1+\frac{\vert\phi_1\vert^2+\vert\phi_2\vert^2}{v^2}\right)-\log\frac{\phi_1\phi_1^*}{v^2}\right]
\]
but gives a  K$\ddot{\rm a}$hler metric formally identical to that in $V_\psi$:
\beqrn h_{1\bar{1}}=\frac{4}{\widetilde{{\cal D}}}-\frac{4 |\phi_1|^2}{v^2{\widetilde{\cal D}^2}}\hspace{2.cm}&&h_{1\bar{2}}=-\frac{4\phi_1^*\phi_2}{v^2{\widetilde{\cal D}}^2}\\
h_{2\bar{1}}=-\frac{4\phi_2^*\phi_1}{v^2{\widetilde{\cal D}}^2}\hspace{2.5cm}&&h_{2\bar{2}}=\frac{4}{{\widetilde{\cal D}}}-\frac{4 |\phi_2|^2}{v^2{\widetilde{\cal D}}^2}\quad ,
\eeqrn
where $\widetilde{{\cal  D}}=1+\frac{\vert \phi_1\vert^2+\vert\phi_2\vert^2}{v^2}$.
The vacuum orbit, however, changes from $\mathbb{S}^3$ to the three-dimensional solid torus $\mathbb{D}^2\times\mathbb{S}^1$, the direct product of the 2D disk times a circle:
\[
\vert\psi_1^{(v)}\vert^2 +\vert\psi_2^{(v)}\vert^2=v^2 \quad \Rightarrow \quad \vert\phi_1^{(v)}\vert^2 -\vert\phi_2^{(v)}\vert^2=v^2.
\]
The explanation of this statement is clear if we write the algebraic equation defining the vacuum orbit in the form $\vert a\vert^2-\vert b\vert^2=1$ where $a=\frac{\phi_1^{(v)}}{\vert v\vert}$ and $b=\frac{\phi^{(v)}_2}{\vert v\vert}$. A point in the solid torus corresponds thus to $(\frac{b}{a},\frac{a}{\vert a\vert})$ since $\frac{b}{a}$ parametrizes the open disk $\mathbb{D}^2$. The closure of $\mathbb{D}^2$ is achieved 
adding the infinity points $\vert a\vert=\infty$, $\vert b\vert=\infty$. It is of note that the solid torus is homeomorphic to the $SU(1,1)$ Lie group formed by the set of matrices $M=\left(\begin{array}{cc} a & b \\ b^* & a^*\end{array}\right)$. 

As a consequence, the $G=U(1)_{\rm gauge}\times SU(2)_{\rm global}$ symmetry exhibited in the reference chart
\[
\left(\begin{array}{c}\psi_1 \\ \psi_2\end{array}\right) \, \rightarrow \, e^{i e \chi(x)}\left(\begin{array}{c}\psi_1 \\ \psi_2\end{array}\right) \quad \quad , \quad \quad \left(\begin{array}{c}\psi_1 \\ \psi_2\end{array}\right) \, \rightarrow \, e^{i\sum_{a=1}^3 \alpha_a \frac{\sigma^a}{2}}\left(\begin{array}{c}\psi_1 \\ \psi_2\end{array}\right)\, ,
\]
where $\sigma^a$ are the Pauli matrices and $\alpha_a$ the parameters of the $SU(2)$ group, collapses to the subgroup $U(1)_{\rm gauge}\times U(1)_{\rm global}$
\[
\left(\begin{array}{c}\phi_1 \\ \phi_2\end{array}\right) \, \rightarrow \, \left(\begin{array}{c}e^{i e \chi(x)}\phi_1 \\ \phi_2\end{array}\right)
\quad \quad , \quad \quad \left(\begin{array}{c}\phi_1 \\ \phi_2\end{array}\right) \, \rightarrow \, e^{i \alpha_3 \frac{\sigma^3}{2}}\left(\begin{array}{c}\phi_1 \\ \phi_2\end{array}\right) \,
\]
in the new chart, whereas the $U(1)_{{\rm gauge}}
$ subgroup only acts on the $\phi_1$ scalar field. Note that this $U(1)_{\rm global}$ is also the subgroup $\left(\begin{array}{cc} e^{i \alpha_3} &  0 \\ 0 & e^{-i\alpha_3}\end{array}\right)$ of $SU(1,1)$.

Regarding the covariant derivatives, in the second chart they are related with their counterparts in the reference chart as follows:
\[
D_\mu\psi_1=-\frac{v^2}{\phi_1^{*2}}\left(\partial_\mu\phi_1^{*}+i e A_\mu \phi_1^{*}\right) \quad , \quad D_\mu\psi_2=-\frac{v\phi_2^*}{\phi_1^{*2}}\left(\partial_\mu\phi_1^*+i e A_\mu \phi_1^{*}\right)+\frac{v}{\phi_1^*}\partial_\mu \phi_2^* \, \, .
\]
We observe again that the field $\phi_2$, coupled to the gauge field only from the K$\ddot{\rm a}$hler potential, is neutral.  The transition functions neatly show that $\phi_2$ is invariant with respect to $U(1)$ gauge transformations.  $\phi_1$, however, carries charge $e$, compelling us to define the covariant derivatives, separately for $\phi_1$ and $\phi_2$, in the compact form
\begin{equation}
D_\mu\phi_p=\partial_\mu\phi_p-i e N_p A_\mu\phi_p \, \, , \, \, p=1,2 \quad , \quad N_1=+1 \, \, , \, \, N_2=0 \, \, . \label{comp}
\end{equation}
Armed with this tool we write the action in identical form in the $V_\phi$-chart to the action in the $V_\psi$-chart introducing a new superpotential to be adjusted by requiring self-duality:
\begin{equation}
S=\int \, dx^4 \, \sqrt{-g}{\cal L} \, , \quad {\cal L}=-\frac{1}{4} {\cal E}(\vert\phi_1\vert,\vert\phi_2,\vert)F_{\mu\nu}F^{\mu\nu}-\frac{1}{2}h_{p\bar{q}}(\phi,\phi^*)D_\mu\phi_pD^\mu\phi_{{q}}^*-\frac{1}{2}\widetilde{W}^2(\phi,\phi^*)\, ,
\end{equation}
where $\phi=(\phi_1,\phi_2)$. Now all the steps starting in formula (\ref{lagrangiano}) and ending in formula (\ref{eleccionP}) are almost identical with some minor differences, e.g. in the behaviour of the $\phi_2$ field. We skip to repeat all this stuff here, but remark that there is an important novelty: the function $P(\psi)$ reduces in the new chart to
\[
\widetilde{P}(\vert\phi_1\vert, \vert\phi_2\vert)=2\frac{\vert\phi_1\vert^2}{\widetilde {\cal{D}}}
\]
because $\phi_2$ is neutral. Therefore self-duality is possible if
\begin{eqnarray}
&& \sqrt{{\cal E}(\vert\phi_1\vert,\vert\phi_2\vert)}\, \widetilde{W}(\vert\phi_1\vert,\vert\phi_2\vert)=e(\widetilde{P}(\vert\phi_1\vert,\vert\phi_2\vert)-v^2) \, \, \Rightarrow  \nonumber\\ && \Rightarrow \widetilde{U}(\vert\phi_1\vert,\vert\phi_2\vert)=\frac{e^2v^4}{2{\cal E}(\vert\phi_1\vert, \vert\phi_2\vert)}\left(\frac{\vert\phi_1\vert^2-\vert\phi_2\vert^2-v^2}{\vert\phi_1\vert^2+\vert\phi_2\vert^2+v^2}\right)^2 \label{potf} \, .
\end{eqnarray}
Thus, we see that self-duality and the change of charts work consistently and, in fact, the potential (\ref{potf}) might be also directly computed applying the transition function to the self-dual potential
working in the $V_\psi$ reference chart. Then, one easily derives the self-duality first-order equations
\begin{eqnarray}
&& F_{12}=\pm\frac{ev^2\sqrt{\gamma}}{{\cal E}(\vert\phi_1\vert,\vert\phi_2\vert)}\left(\frac{\vert\phi_1\vert^2-\vert\phi_2\vert^2-v^2}{\vert\phi_1\vert^2+\vert\phi_2\vert^2+v^2}\right) \label{sd1}\\&& D_1\phi_p\pm i\sqrt{\gamma}D^2\phi_p=0 \, \, , \, \, p=1,2\label{sd2}
\end{eqnarray}
We omit that calculation, but the proof that the solutions of (\ref{sd1})-(\ref{sd2}) have null the planar components of the energy momentum tensor, $T_{ij}=0$, runs in parallel to the same demonstration for self-dual cosmic strings of the first species.
Finally, we must solve together with (\ref{sd1})-(\ref{sd2}) the Einstein equation
\begin{equation}
R^{(\gamma)}=16 \pi G  T_{tt} \quad \label{einstein5}
\end{equation}
As it is described in the Subsection 3.3 for the reference chart, the equations (\ref{sd2}) in the conformal gauge, $\gamma_{ij}=e^{-4 \widetilde{V}(x^1,x^2)}\delta_{ij}$, become $D_1\phi_p\pm D_2\phi_p=0$, i.e., are identical to the akin equations in the Euclidean plane $\bar{\partial}_A\phi_p=0$.  Therefore, for the field $\phi_2(x_1,x_2)$, (\ref{sd2}) is the Cauchy-Riemann equation in this conformal gauge. The $\phi_2$-solutions are accordingly holomorphic/antiholomorphic functions $\phi_2=f(x^1\pm ix^2)$, Together with the finite energy conditions we shall keep only constant functions as bona fide solutions. The boundary condition $\lim_{z\to +\infty}\phi_2(z)=w_2\in\mathbb{C}$ then sets $\phi_2(z)=w_2$ as the solutions throughout the whole plane. To reach the vacuum orbit at infinity the limit of the other field at the boundary $\lim_{z\to +\infty}\phi_1(z)=w_1\in\mathbb{C}$ is forced to satisfy the equation $\vert w_1\vert^2-\vert w_2\vert^2=v^2$. We stress that the string tension of self-dual solutions in this chart is also $\mu=2\pi \vert n\vert v^2$ like the tension of cosmic strings in the Reference chart. The tension of this species of cosmic strings fits with the tension of non-gravitating strings in the analogous chart described in \cite{nosotros15} if we identify $v^2=\rho^2$ where $\rho$ is the parameter introduced in \cite{nosotros15}.

The vortex equation (\ref{sd1}) may be rewritten {\it \` a la} Jaffe-Taubes \cite{jatau80}
\begin{equation}
\bigtriangleup \log \vert \phi_1\vert=e^2 v^2 e^{-4\widetilde{V}}\frac{\vert\phi_1(z,\bar{z})\vert^2-\vert w_2\vert^2-v^2}{\vert\phi_1(z,\bar{z})\vert^2+\vert w_2\vert^2+v^2}+2\pi \sum_{s=1}^n\delta^{(2)}(z-z_s)\, . \label{vortex}
\end{equation}
The dimension of the moduli space of solutions of this equation (\ref{vortex}) is $2n+2$ due to the freedom of moving the centers of the cosmic strings $z_s$ and the freedom of choosing a point in $\mathbb{D}^2\times\mathbb{S}^1$ by changing $w_2$.  In particular, the choice $\phi_1^{(v)}=v$, $\phi_2^{(v)}=0$ selects as the self-dual cosmic strings in this chart identical solutions to the self-dual cosmic strings in the $\mathbb{CP}^1$-sigma model.

It remains to solve the Einstein equation (\ref{einstein5}). In the conformal gauge  this equation becomes
\begin{equation}
\bigtriangleup \Big[\widetilde{V}-4\pi Gv^2\left(\log\widetilde{\cal D}- \frac{1}{2}\log \frac{\vert\phi_1(z,\bar{z})\vert^2}{\prod_{s=1}^n \vert z-z_s\vert^2}\right)\Big]=0 \label{harm}
\end{equation}
Therefore, the quantity between brackets is harmonic and the solution for the metric is:
\begin{equation}
\widetilde{V}(z,\bar{z})=4\pi Gv^2\left(\log\frac{v^2+\vert\phi_1(z,\bar{z})\vert^2+\vert w_2\vert^2}{v^2}- \frac{1}{2}\log \frac{\vert\phi_1(z,\bar{z})\vert^2}{\prod_{s=1}^n \vert z-z_s\vert^2}\right)+{\rm constant}.
\end{equation}
The search for cylindrically symmetric cosmic strings of the second species proceeds along similar lines to the task performed on the first species. Thus, we stop here to avoid unnecessary repetitions. Nevertheless, we emphasize the main features of this second species of self-dual cosmic strings:

\begin{enumerate}

\item Like the cosmic strings of the first species, the cosmic strings of the second species are concentrated in the plane perpendicular to the string. Contrarily to the first species they take values in the subset of $V_\phi$, instead of $V_\psi$, comprised between the point ($\phi_1=0$, $\phi_2=w_2$) and the vacuum orbit $\mathbb{D}^2\times \mathcal{S}^1$. This is a region of $\mathbb{CP}^2$ disjoint to the region in which cosmic strings of the first species are evaluated. In particular, if $w_2=0$ the cosmic strings are mapped into the spherical cap of $\mathbb{CP}^1$ complementary to the spherical cap in $V_\psi$ where the cosmic strings of the first species live.

\item Solutions of both species can be described using any chart. Nevertheless, at the center of a cosmic string of the first species $\psi_1=0$, so that $\phi_1=\infty$ and the string looks as a singular configuration from the point of view of $V_\phi$. The same is true for the second species as seen from the reference chart. Thus, the most natural treatment is to use $V_\psi$ and $V_\phi$ for, respectively, first and second species.

\item Since there is a potential energy for $\mathbb{CP}^2$ fields, solutions taking values in different bounded regions of $\mathbb{CP}^2$ come from different effective dynamics, which can therefore lead to different physical properties. Nevertheless, cosmic strings of both species have in our model the same energy per unit length.

\item One might think of a configuration encompassing a cosmic string of the first species plus another string of the second species, both of voticity $n$. Since the total string tension adds to $4\pi \vert n\vert v^2$ and this is precisely the degree
of a map from $\mathbb{S}^2$ to $\mathbb{CP}^N$ that characterizes the $\mathbb{CP}^N$-instantons this combination of two strings of different species could be interpreted as a partonic (or meronic) structure of the instanton. A caveat: even BPS vortices suffer a low energy dynamics, see e.g, \cite{WJ}
and references quoted therein. Thus, the double cosmic string configuration is not solution of the static field equations. Only in the limit $G=0$, $e^2=0$ the instanton partonic structure can survive.
\end{enumerate}

% \begin{enumerate}

% \item Like the cosmic strings of the first species the cosmic strings of the second species are concentrated in the plane perpendicular to the string. Contrarilly to the first species they takes value in the subset of $V_\phi$, instead of $V_\psi$, comprised between the point $\phi_1(0,0)=0$,$\phi_2(0,0)=w_2$ and the vacuum orbit $D\times S^1$.If $w_2=0$ the cosmic strings are mapped in the spherical cap of $\mathbb{CP}^1$ complementary to the spherical cap in $V_\psi$ where the cosmic strings of the first species live.

% \item One might think of a configuration encompassing a cosmic string of the first species plus another string of the second species, both of voticity $n$. Since the total string tension adds to $4\pi \vert n\vert v^2$ and this is precisely the dgree
% of a map from $\mathbb{S}^2$ to $\mathbb{CP}^N$ this combination of two strings of different species could be interpreted as a partonic (or meronic) structure of the instanto. A caveat: even BPS vortices suffer a low energy dynamics, see e.g, \cite{WJ}
% and references quoted therein. Thus, the double cosmic string configuration is not solution of the static field equations. Only in the limit $G=0$, $e^2=0$ the instanton partonic structure can survive.
% \end{enumerate}}

\section{Final comments}
The main theme of this investigation has been the cosmic strings, Type I, Type II, and self-dual, existing in Abelian gauged non-linear $\mathbb{CP}^2$-sigma model coupled to gravity. Thus, we address a concrete case in the general framework of research about topological defects in gravitational scenarios. This is a fertile subject with many branches, some more theoretical, see for instance a former work \cite{nosotros20} centred around kinks interacting with the gravitational field created by themselves in Jackiw-Teitelboim $(1+1)$- dimensional pseudo-Riemannian universes, and some more directly connected to realistic cosmological problems, like in the present case. Namely, we have studied topological defects and the gravitational fields created by them in the context of  Abelian gauged non-linear sigma models coupled to Einstein gravity in $(3+1)$-dimensional pseudo-Riemannian universes.

A first remark about the construction of the cosmic strings achieved in this work 
%within the framework of Abelian gauged non-linear $\mathbb{CP}^2$-sigma models coupled to gravity%
is the observation that trading the target manifold $\mathbb{CP}^2$ by $\mathbb{CP}^N$ the same structure arises. There is a first species of cosmic strings living in the reference chart, but identical cosmic strings of the second species appear in the remaining $N$ charts that, together with the reference chart, form a minimum atlas of $\mathbb{CP}^N$.

It is also of note that, in comparison with the non-gravitating solitonic strings of Reference \cite{nosotros15}, here the parameter space is reduced for simplicity. From $\rho$, the radius of the sphere which determines the Fubini-Study metric, and $a$, which sets the scale of the gauge symmetry breaking, in \cite{nosotros15}, in this work there is a reduction to a single parameter $v$ which determines both the K\"{a}hler metric and the vacuum expectation value of the scalar fields. As a consequence, the differences between the cosmic strings of the two species are less evident, but the parameter space used in \cite{nosotros15} could be easily translated to the gravitating case dealt with in the present work.

Another possible route to extend this work is the choice of more general dielectric functions. In Reference \cite{Bazeia20} Bazeia et al. proposed an interesting family of dielectric functions susceptible of supporting self-dual topological vortices in shrewd generalizations of the Abelian Higgs model. The main novelty is the existence of vacuum orbits formed by several connected components. The challenge is to promote these structures to non-linear gauged sigma models coupled to gravity.

Finally, we comment about the possibility of introducing another K\"{a}hler metrics different from the Fubini-Study metric. The existence of self-dual cosmic strings when the target manifold is endowed with general K\"{a}hler metrics is
%guaranteed% 
to be expected because these metrics permit the extension of Abelian gauged non-linear sigma models to their ${\cal N}=2$ supersymmetric partners \cite{Intriligator}.
\section*{Acknowledgements}
The authors acknowledge the Junta de Castilla y Le\'on for financial support under grants BU229P18 and SA067G19.

\end{document}